\documentclass[fleqn,usenatbib]{mnras}

\usepackage{newtxtext,newtxmath}

\usepackage[T1]{fontenc}

\DeclareRobustCommand{\VAN}[3]{#2}
\let\VANthebibliography\thebibliography
\def\thebibliography{\DeclareRobustCommand{\VAN}[3]{##3}\VANthebibliography}

\usepackage{graphicx}	
\usepackage{amsmath}	

\title[NIGHT: Surveying Helium in Exoplanets]{NIGHT: a compact, near-infrared, high-resolution spectrograph to survey helium in exoplanet systems}

\author[C. Farret Jentink et al.]{C. Farret Jentink,$^{1}$\thanks{E-mail: casper.farret@unige.ch}
        V. Bourrier,$^{1}$
        C. Lovis,$^{1}$
        R. Allart,$^{2}$\thanks{Trottier Postdoctoral Fellow}
        B. Chazelas,$^{1}$
        M. Lendl,$^{1}$
        X. Dumusque,$^{1}$
        F. Pepe$^{1}$
\\
$^{1}$Observatoire Astronomique de l’Universit\'e de Gen\`eve, Chemin Pegasi 51b, 1290 Versoix, Switzerland\\
$^{2}$D\'epartement de Physique, Institut Trottier de Recherche sur les Exoplan\`etes, Universit\'e de Montr\'eal, Montr\'eal, Qu\'ebec, H3T 1J4, Canada\\
}

\date{Accepted 2023 October 19. Received 2023 October 18; in original form 2023 June 14}
\pubyear{2023}

\begin{document}
\label{firstpage}
\pagerange{\pageref{firstpage}--\pageref{lastpage}}
\maketitle

\begin{abstract}
Among highly irradiated exoplanets, some have been found to undergo significant hydrodynamic expansion traced by atmospheric escape. To better understand these processes in the context of planetary evolution, we propose NIGHT (the Near-Infrared Gatherer of Helium Transits). NIGHT is a high-resolution spectrograph dedicated to surveying and temporally monitoring He I triplet absorption at 1083nm in stellar and planetary atmospheres. In this paper, we outline our scientific objectives, requirements, and cost-efficient design. Our simulations, based on previous detections and modelling using the current exoplanet population, determine our requirements and survey targets. With a spectral resolution of 70,000 on a 2-meter telescope, NIGHT can accurately resolve the helium triplet and detect 1\% peak absorption in 118 known exoplanets in a single transit. Additionally, it can search for three-sigma temporal variations of 0.4\% in 66 exoplanets in-between two transits. These are conservative estimates considering the ongoing detections of transiting planets amenable to atmospheric characterisation. We find that instrumental stability at 40m/s, less stringent than for radial velocity monitoring, is sufficient for transmission spectroscopy in He I. As such, NIGHT can utilize mostly off-the-shelf components, ensuring cost-efficiency. A fibre-fed system allows for flexibility as a visitor instrument on a variety of telescopes, making it ideal for follow-up observations after JWST or ground-based detections. Over a few years of surveying, NIGHT could offer detailed insights into the mechanisms shaping the hot Neptune desert and close-in planet population by significantly expanding the statistical sample of planets with known evaporating atmospheres. First light is expected in 2024. 
\end{abstract}



\section{Introduction}

Over recent years, several exoplanets have been found to undergo atmospheric escape -- losing their volatile upper atmosphere to space due to strong heating and subsequent hydrodynamic escape \citep{vidal2003extended, lammer2003}. The dominant source of heating is XUV (X-ray and extreme UV) irradiation from the host star \citep{owen2019}, but core-powered mass loss could also play a role for small gas-rich planets \citep{gupta2019, modirrousta2022three}. The importance of atmospheric escape in a planet’s evolution is not fully understood, in particular because there is a strong lack of observations to support models \citep{salz2016,mordasini2020}. Mass-loss rates estimated to date suggest that a significant fraction of low-mass planet's atmospheres can be lost to space, especially for Neptune-sized planets \citep{bourriergj436b, owen2018}. Consequently, hydrodynamic escape is believed to play a vital role in the formation of the hot Neptune Desert, eroding these worlds into their bare rocky cores \citep{etangs2004atmospheric, bourriergj436b, owen2018, attia2021, koskinen2022mass}. 

The process of atmospheric evaporation was first observed through excess absorption in the Lyman-$\alpha$ line by \cite{vidal2003extended} in transits of HD\;209458\;b. The depth of the absorption signature was found to correspond to an atmospheric layer extending beyond the Roche lobe of the planet, thus leading to the conclusion of an escaping atmosphere. Lyman-$\alpha$ proved to be a good tracer for hydrodynamic escape since upper atmospheres are mostly composed of light elements like hydrogen and helium. Since this first detection, other planets have been found to display strong excess absorption in Lyman-$\alpha$, e.g. HD\;189733\;b \citep{des2010evaporation, lecavelier2012, bourrierhd189, bourrierhd189_2}, GJ\;436\;b \citep{kulow2014lyalpha, ehrenreich2015giant, lavie2017, bourriergj436b, santos2019} or GJ3470 b \citep{bourriergj3470}. However, using Lyman-$\alpha$ as a tracer for atmospheric escape comes with a few caveats. Firstly, the Hubble space telescope is needed as it is currently the only UV observatory with enough sensitivity, spectral coverage, and spectral resolution to observe this line. Observations are limited to nearby planetary systems due to strong absorption by the interstellar medium, Earth's hydrogen geocorona, and Hubble’s own optics. A possible successor like LUVOIR or HABEX \citep{luvoir2019luvoir, gaudi2020habitable} will likely only be launched in a couple of decades. 

   \begin{figure*}
   \centering
   \includegraphics[width=0.49\textwidth]{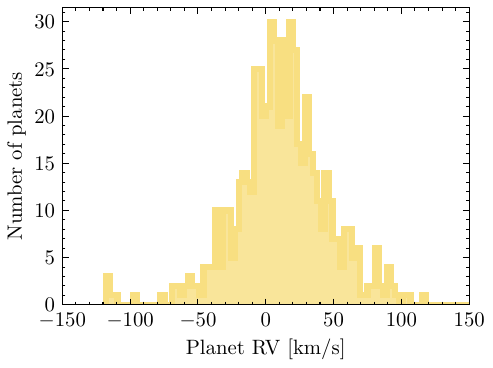}
   \includegraphics[width=0.49\textwidth]{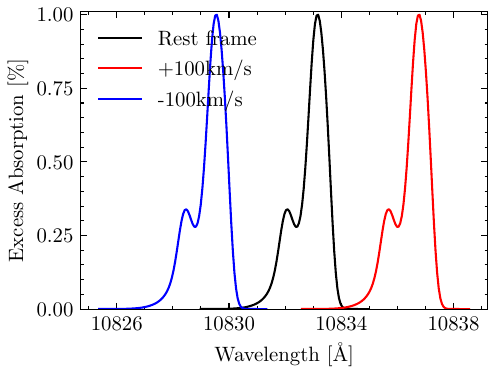}
   \caption{Line-of-sight velocities histogram of known exoplanets and the helium absorption signature both in a 0km/s rest frame and positive and negative 100km/s shifted frame. The model of the helium absorption signature was taken from \citet{allart2019high} and normalised at 1\% for visualisation purposes.}
              \label{fig:LOS}%
    \end{figure*}

To continue the study of extended and evaporating atmospheres, an alternative tracer at longer wavelengths and/or observable from the ground is needed. This tracer was found in a triplet of the metastable helium state around 10,833 \AA\ in the near-infrared \citep{seager2000theoretical, oklopcic2017}. It was observed in WASP-107\;b \citep{spake2018helium, allart2019high, kirk2020confirmation, spake2021}, HAT-P-11\;b \citep{allart2018spectrally}, WASP-69\;b \citep{nortmann2018}, HD\;189733\;b \citep{salz2018,zhang2022more}, and more recently in WASP-52\;b \citep{kirk2022keck}, and TOI-560\;b \citep{zhang2022detection}, among others. This helium triplet proves to be an excellent tracer of atmospheric escape, being much less sensitive to absorption from the interstellar medium, observable from the ground, and sensitive enough to trace the rarefied gas in extended atmospheres. 

To better understand the relation between atmospheric escape and planetary evolution, a large sample of helium observations is now required to:
 \begin{itemize}
     \item study the structure of upper atmospheres and planetary winds and the dependence of mass loss on stellar and planetary properties;
     \item investigate the temporal variability of atmospheric structure and mass loss;
     \item determine the role of atmospheric escape in planetary evolution.
 \end{itemize}
 
Several high-resolution (HR) spectrographs dedicated to near-infrared observations and able to resolve the helium triplet have been developed in recent years: CARMENES at the Calar Alto 3.5 \citep{quirrenbach2018carmenes}, NIRSPEC at Keck \citep{mclean1998design}, SPIRou at CFHT \citep{artigau2014spirou}, GIANO-B at the TNG \citep{origlia2014}, and NIRPS at the ESO 3.6 \citep{wildi2017nirps, bouchy2017near}, among others. However, all these instruments are placed at highly competitive, 3.5+ meter class telescopes. To allow for a significant increase in helium detections and long-term monitoring of exoplanet atmospheres, we propose the Near-Infrared Gatherer of Helium Transits (NIGHT), a high-resolution, narrowband spectrograph solely dedicated to measuring helium absorption and spectrally resolving the lineshape during exoplanet transits. A previous survey-type instrument for exoplanet helium absorption was presented in \citet{vissapragada2020}, utilising ultra-narrowband photometry to detect these atmospheres. However, their instrument was not able to spectrally resolve the triplet -- which is essential to the studies we want to conduct.

Because of its specific science case, NIGHT can be optimised in efficiency, allowing for a survey on a smaller telescope than the aforementioned instruments are placed on. Our initial aim is to place NIGHT as a visitor instrument on a 2-meter class telescope -- while keeping in mind possible switches to other (larger) telescopes later in life. Next to being able to optimise for efficiency, the narrow passband of NIGHT allows the use of mostly off-the-shelf components, reducing instrument development time, complexity, and overall instrument cost. Altogether, the specialised science case opens up opportunities and simplifications in design that have not been exploited before in HR spectrographs for exoplanetary science. In the following chapters we first present the science requirements, and simulations leading up to technical requirements, followed by the preliminary targets for our survey and instrument design. 

\section{Science and Technical Requirements}
\label{sec:reqs}
The top-level science requirements of NIGHT have been defined as the following:
\begin{enumerate}
    \item Observe helium transits of exoplanets with sufficient sensitivity, temporal, and spectral sampling to probe the dynamics of the extended atmosphere and its variability;
    \begin{enumerate}  
        \item Detect 1\% peak excess helium absorption at 5 sigma for at least 100 planets;
        \item Detect 0.4\% temporal variability in peak absorption depth between two transits at 3 sigma for at least 50 planets (similar to the detection in \citet{allart2018spectrally});
    \end{enumerate}
    \item Have two channels, one science, and one on-sky;
\end{enumerate}

Furthermore, we would like to note that it is our aim to keep the instrument cost-efficient and compact. To define the technical requirements of NIGHT, we used past detections of the He I triplet with HR spectrographs and the currently known exoplanet population as references to estimate observable signals. Based on these estimates we could then define the instrument requirements allowing us to fulfill the science requirements. We finally explored technical designs to fulfill the requirements. 

\subsection{Wavelength band}
For the wavelength coverage, it is important to accommodate the full width of the helium signature and the surrounding continuum, accounting for its line-of-sight (LOS) velocity shift and the necessity to access wavelength calibration source lines. In Figure~\ref{fig:LOS} we show a histogram of LOS velocities for all exoplanets known at the moment of writing, as taken from the NASA Exoplanet Catalog and The Extrasolar Planets Encyclopaedia \citep{akeson2013nasa, schneider1996extrasolar}. The LOS velocity has been computed by the addition of the stellar system velocity and the planet ingress velocity. As can be seen, almost all velocities fall within a -100km/s to +100km/s range. Figure~\ref{fig:LOS}b further shows the modeled helium signature in both rest- and shifted frame(s). The model of the helium absorption signature was taken from \citet{allart2019high}. Based on the expected LOS shifts, we should have minimal spectral coverage of 10,826 -- 10,838 \AA\ for NIGHT. Taking into account a maximum barycentric velocity shift of 30km/s this widens to a band of $\sim$10,825 -- 10,839 \AA. Adjusting the wavelength range for science alone is not enough. High-precision, stabilised spectrographs need to be wavelength calibrated regularly as environmental changes like pressure or temperature can alter the spectral response, both on short and long timescales \citep{pepe2014instrumentation}. To calibrate the wavelength scale of NIGHT, we propose to use both:
\begin{enumerate}
    \item A Uranium-Neon (Ur-Ne) hollow cathode lamp;
    \item Telluric absorption and emission lines (if strong enough, given their variability).
\end{enumerate}

\begin{figure}
   \centering
   \includegraphics[width=\columnwidth]{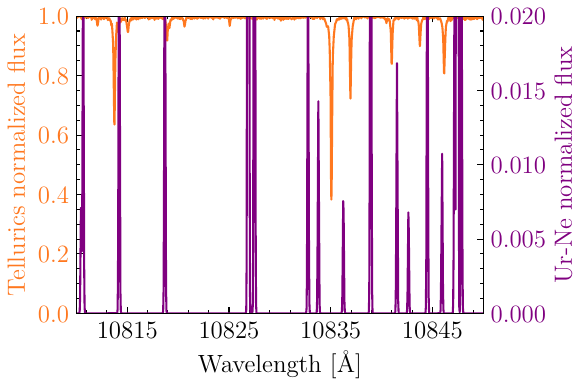}
      \caption{Calibration source lines from 10810\AA\ to 10845\AA. In purple, we find the Ur-Ne emission lines from the \citet{redman2012high} catalog, and in orange the telluric absorption spectrum from \citet{wallacelivingston2003}.
              }
         \label{fig:calib}
   \end{figure}

Both have proven to be good calibration sources, providing precision better than 50m/s \citep{redman2011infrared, redman2012high, figueira2010evaluating}. The corresponding spectra of the telluric lines and Ur-Ne emission lamp can be found in Figure~\ref{fig:calib}. To have a decent calibration, we estimate a minimum of 8 lines is needed. As such, to accommodate for at least 8 Ur-Ne and 8 telluric absorption lines, we propose a spectral coverage of 10,810 to 10,850 \AA\ for NIGHT. If atmospheric conditions allow, we also propose to use telluric OH emission lines for wavelength calibration.

\subsection{Spectral resolution}
\begin{figure}
   \centering
   \includegraphics[width=\columnwidth]{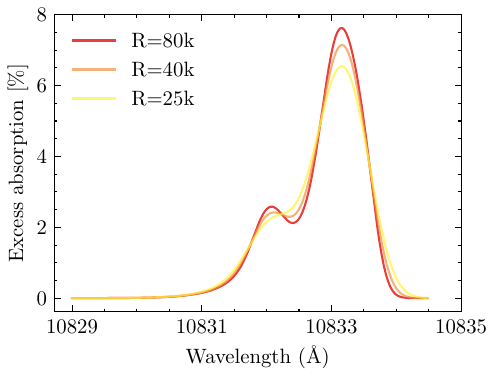}
      \caption{A model of the He I triplet from \citet{allart2019high} at three different spectral resolutions, R = 80k, 40k, and 25k. The convolutions illustrate that spectral resolutions of R = 40k and 25k do not fully resolve the lineshape.
              }
         \label{fig:spec_convol}
   \end{figure}

The spectral resolution of a spectrograph impacts to which degree one can resolve individual features and/or lines in the spectrum, but also the accuracy with which one can determine the depth or strength of these signatures. The spectral resolution is often denoted in terms of resolving power $R$:
\begin{equation}
    \label{eq:resolving}
    R = \frac{\lambda}{\Delta\lambda},
\end{equation}
where $\Delta\lambda$ is the smallest wavelength scale that can be distinguished at wavelength $\lambda$. For the He I triplet there are two aspects to consider: a) whether we can distinguish the two separate absorption features created by the three lines (the two strongest, reddest lines are naturally blended due to thermal Doppler broadening), and b) the accuracy of the measured peak absorption for the strongest peak. To calculate the Doppler broadening for a heated helium-dominated thermosphere we can use Equation~\ref{eq:therm}:
\begin{equation}
    \label{eq:therm}
    \Delta\lambda = \lambda_0 \cdot 2 \sqrt{2\mathrm{ln}2\frac{kT}{m_0 c^2}}
\end{equation}
For a thermosphere with an equilibrium temperature of 5000K \citep{salz2016}, and $\lambda_0 = 10,833 \AA$, this returns a Doppler broadening of 0.275 \AA. Following Equation~\ref{eq:resolving} and \ref{eq:therm}, this implies a minimal resolving power $R\sim 40,000$ to resolve the lineshape. Depending on the line-depth of the He I lines, a higher spectral resolution might however be preferred. To demonstrate this effect, we show a model of a detection made by \citet{allart2019high}, with the CARMENES spectrograph at R=80k in Figure~\ref{fig:spec_convol}. The 80k resolution model is convolved at spectral resolutions of 25k, and 40k, to illustrate what would have been measured at lower spectral resolutions.
We can see that the three models indicate different maximum absorption depths. Although we would agree that at resolutions of R=25k and 40k, one can resolve the shape of the triplet to some extent, a higher spectral resolution is required to fully resolve the two bumps at their typical thermal broadening. Spectral resolutions that fully resolve the shape of the He I triplet will also allow for improved accuracy when temporally resolving changes in the absorption depth between transits because we can be more accurate on the peak absorption. It is important to note here that for now we disregard the signal-to-noise ratio (SNR) on the spectrum. The precision on the peak absorption depth will degrade with finer sampling as the flux per pixel drops, decreasing the SNR.

\subsection{Spectral sampling}
Sampling impacts the spectrum in a similar fashion but inherently impacts the maximum SNR on the spectral bins. A higher SNR implies smaller amplitude signals can be distinguished -- in our science case essential to potentially reveal small amplitude variations in the absorption depth. As such, we would like to minimise the number of pixels to increase SNR in single pixels, without losing spectral information. The Nyquist-Shannon theorem determines the minimum sampling size of pixels – all spectral information of a bandwidth-limited signal is conserved if at least 2 pixels sample a single resolution element \citep{shannon1949communication}. However, most modern-day high-resolution spectrographs opt for 3 pixels per resolution element to assure the shape of narrow spectral features is guided by the instrumental point spread function (PSF) and not pixel sampling, aiding in data reduction. Also, narrow spectral lines are of Gaussian-like shape, implying they are not bandwidth-limited and there is no sharp 2.0-pixel sampling limit.

   \begin{figure}
   \centering
   \includegraphics[width=\columnwidth]{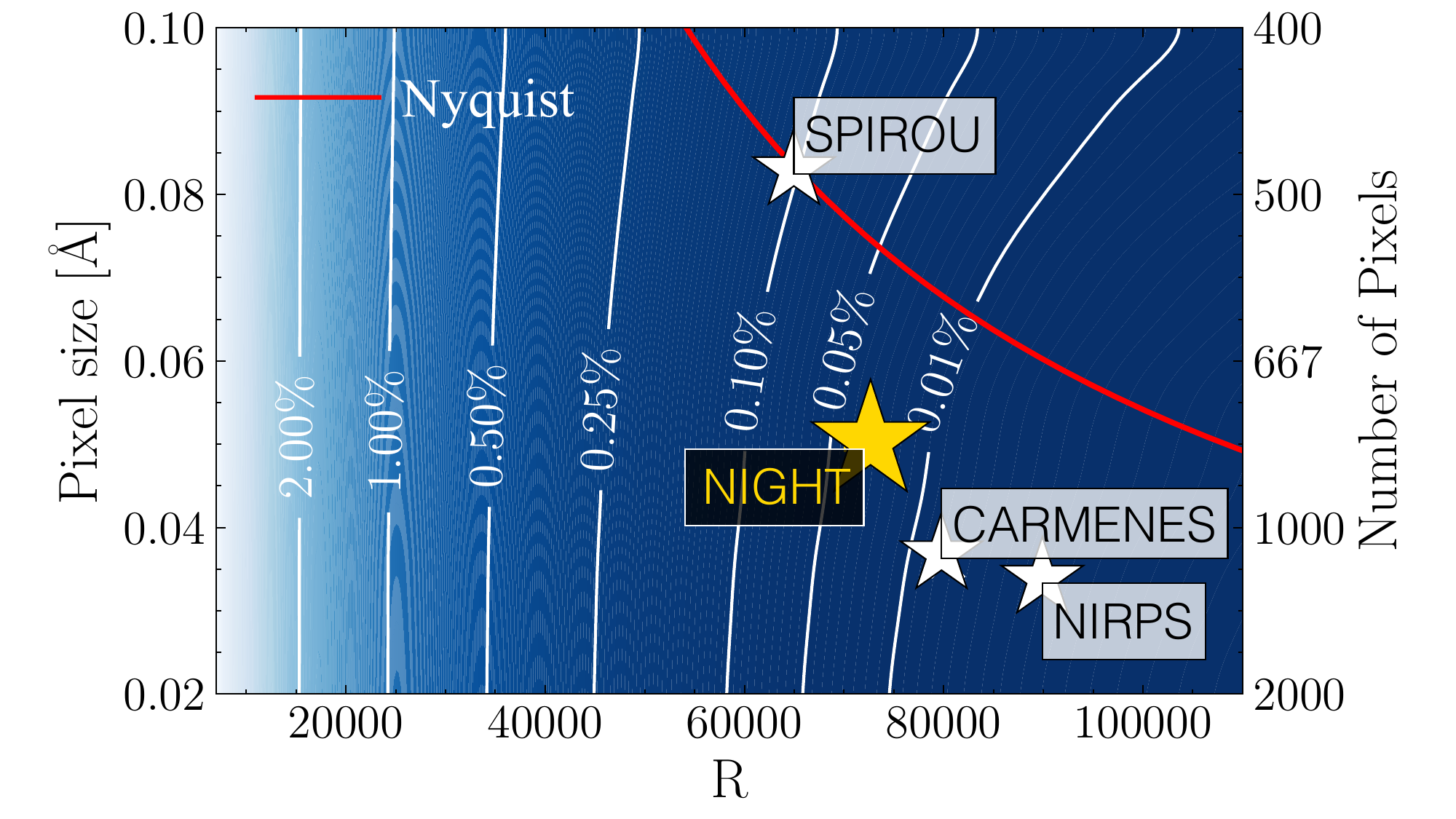}
      \caption{Peak absorption offset in the percentage of total flux systematically induced by not being able to resolve the triplet. The horizontal axis indicates the spectral resolution $R$, the left vertical axis the pixel size in units of wavelength, and the right vertical axis the total number of pixels required in the 10810~--10850 \AA\ band. The red line shows the Nyquist threshold where 2 pixels sample each resolution element $\Delta\lambda$. Above this line, the spectrum is undersampled. As expected, for higher spectral resolutions, the offset decreases. Various markers are plotted for existing HR spectrographs at their corresponding spectral resolution and pixel size. The gold marker denotes the proposed spectral resolution and pixel size for the NIGHT spectrograph.
              }
         \label{fig:res_sampl}
   \end{figure}

To determine the ideal pixel sampling and spectral resolution for NIGHT, we convolved the model of the helium signature from \citet{allart2019high} over a range of PSF widths and sampled it over a range of pixel sizes. We then studied the systematic loss of absorption signal at the peak absorption signature between the original model and the convolved, resampled signatures (see Fig.~\ref{fig:res_sampl}). We determined the level of peak absorption by fitting a Gaussian to the strongest peak.

From the contours in Figure~\ref{fig:res_sampl}, we find that spectral resolutions above R=70,000 do not introduce significant absorption losses. This is expected since the model signature is already significantly broadened by thermodynamic effects within the planet's atmosphere. Pushing the spectral resolution to higher values does not improve the accuracy and would decrease SNR for finer pixel sampling as we keep the sampling per spectral bin fixed. Furthermore, we see that below the Nyquist threshold, finer sampling does not significantly increase the accuracy. At R=70,000, we find an offset at a level of 0.05\% at Nyquist sampling -- just a fraction of any to-be-expected photon noise. As such, we propose a spectral resolution of 70,000 -- 75,000 for NIGHT to fully resolve the He I triplet at its typical broadening. As for the sampling, Nyquist sampling would be sufficient for the helium triplet but we propose to increase to >2.5 pixels per resolution element ($\sim$ 0.055 \AA\ per pixel) to sufficiently sample narrow spectral lines.

\subsection{Wavelength Precision and Stability}
\label{sec:wlstab}

    \begin{figure*}
    \centering
    \includegraphics[width=0.52\textwidth]{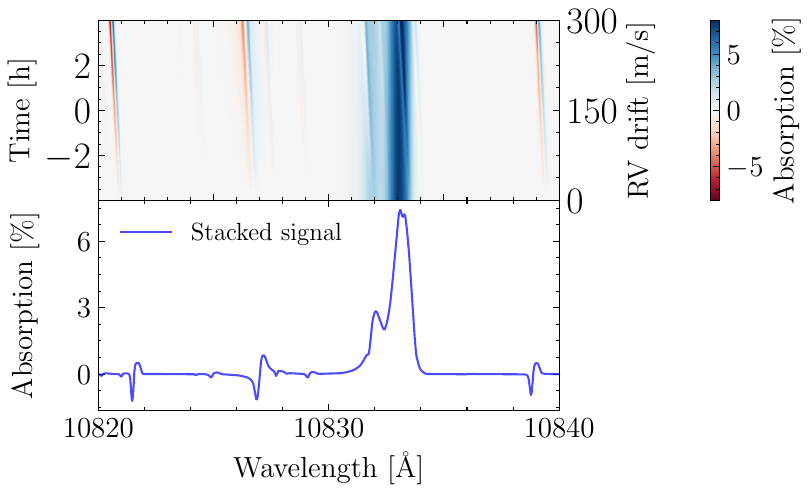}
    \includegraphics[width=0.46\textwidth]{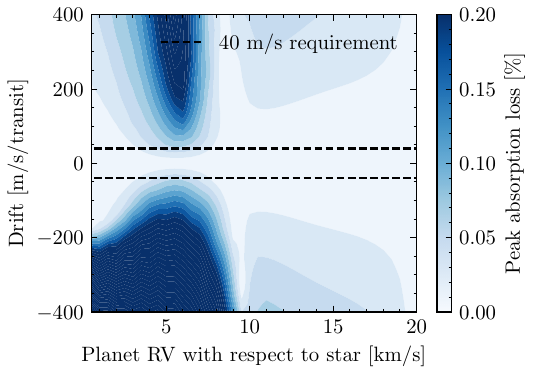}
    \caption{On the left-hand side we show a simulated transit in the planetary rest frame. In the upper panel, we find a river diagram of 50 master-out corrected spectra. The transit takes 8 hours and the planet has an RV of -5km/s w.r.t. the star at ingress. These values were taken to amplify the effect of drift (as seen in the right panel), as a worst-case scenario. We applied an uncorrected linear instrumental drift from 0m/s up to 300m/s toward the end of the transit. This uncorrected drift introduces p-cygni profiles in stellar absorption lines that, as a result of the uncorrected drift, are not divided out completely. The lower panel shows the stacked absorption spectra. On the right-hand side, we find the offset at the peak absorption introduced by a linear drift over a full transit for a range of planet RVs at ingress.}
              \label{fig:drift}%
    \end{figure*}

HR spectrographs for exoplanetary science have typically been developed with mass determination through a radial velocity (RV) signature of planets in mind. This requires a high RV precision that in turn calls for high instrumental stability, especially when aiming at detecting tiny signals introduced by low-mass, long-period planets. For example, the ELODIE spectrograph \citep{baranne1996elodie}, which was used to detect the first exoplanet around a Sun-like star, 51Peg b \citep{mayor1995jupiter}, was able to calibrate wavelengths to a precision of 15 m s$^{-1}$. This allowed for the detection of hot Jupiters, massive planets close to their host stars. The latest generation of instruments like ESPRESSO \citep{pepe2014espresso,pepe2021} aims at detecting rocky planets in the habitable zone of solar-type stars, requiring an RV precision in the order of 0.1 m s$^{-1}$.
For atmospheric characterisation of planets through transits, no such extreme RV precision or instrument stability as for ESPRESSO is required. 

   \begin{figure}
   \centering
   \includegraphics[width=\columnwidth]{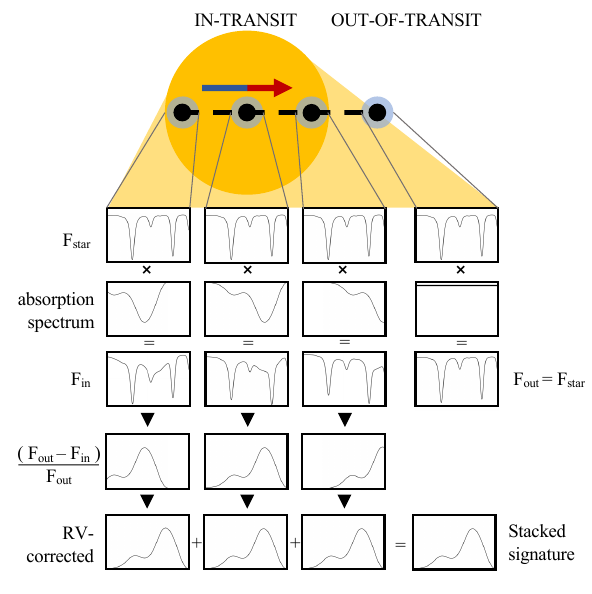}
      \caption{A visual representation of the observations performed during exoplanet transit spectroscopy and how an excess absorption signature for He I is retrieved. Multiple measurements are performed while the planet is in transit. The measured spectrum ($\mathrm{F_{in}}$) is a multiplication of the stellar spectrum ($\mathrm{F_{star}}$) and the planet's absorption spectrum. Similarly, a master spectrum ($\mathrm{F_{out}}$) is measured from a stack of out-of-transit stellar spectra. The retrieved absorption spectra are computed with Equation~\ref{eq:trans}. After an RV correction (induced by the fact that the planet changes in LOS velocity during transit), the retrieved spectra can be stacked.}
         \label{fig:transit_sim_fig}
   \end{figure}

To determine the required RV precision and stability for NIGHT, we first need to understand how it will perform its measurements. In principle, two measurements of the stellar spectrum are required; one when the planet is transiting; a second when it is out of transit. An out-of-transit measurement consists of multiple, stacked exposures to increase the signal-to-noise ratio (SNR), creating a master stellar spectrum. By subtracting an in-transit spectrum from the master stellar spectrum and normalising over the master spectrum, an absorption spectrum of the exoplanet’s atmosphere is generated following Equation~\ref{eq:trans}:
\begin{equation}
    A_{planet}[\%] = \frac{F_{out}-F_{in}}{F_{out}}.
    \label{eq:trans}
\end{equation}
The independent in-transit measurements are all individually compared to the master stellar spectrum and are only stacked after having been corrected for the changing orbital velocity of the planet. Figure~\ref{fig:transit_sim_fig} shows an illustration of this process.

The requirements on precision and stability can be derived from how precise the in-transit and out-of-transit spectra’s wavelength solutions need to be calibrated to still lead to an accurate retrieval of the absorption spectrum of the planet's atmosphere. In Figure~\ref{fig:drift}, we show a simulated measurement of a full transit. Our simulations show that spurious features are introduced in the final absorption signature as a result of uncalibrated instrumental drift. If the stellar continuum would have been flat in our wavelength band, we would have not seen any artifacts. However, stellar lines are present in both measurements. To subtract these from the retrieved signature, they need to be well aligned. An uncalibrated radial velocity drift over time can shift stellar lines from their original location – leading to inaccurate subtraction and spurious features in the retrieved signature, like p-cygni profiles. This in return will lead to inaccuracies in the shape of the He I absorption feature and its absorption depth. To quantify this effect, we ran retrievals, introducing a linear instrumental RV drift over a transit. The resulting retrievals were compared to an ideal case, un-drifted scenario to put a requirement on wavelength calibration precision, and with that on wavelength stability. Similarly, we applied the same method for a hypothetical resolution change of the instrument between in- and out-of-transit spectra. 

In Figure~\ref{fig:drift} we show the results of our RV stability simulations. We can see that for negative drifts, the peak absorption offset grows faster than for positive drifts. This is caused by the strong stellar silicon line moving towards the peak of the He I triplet. We also see that there is a strong dependence on the orbital velocity of the planet. We need to be most stringent on RV drift around an ingress velocity of 5km/s. For higher velocities, the stellar-line-induced p-cygni profiles move significantly enough over the spectrum to mostly cancel out when stacked afterward. For lower velocities, the stellar lines move less and create smaller offsets. Given the fact that we will be looking for temporal variations at the level of 0.4\% in the helium absorption signature's depth (see Section~\ref{sec:reqs}) for a wide variety of planets, it is important to keep inaccuracies as a result of limited instrument stability and drift well below this level. We decided to put the threshold of acceptable inaccuracy as caused by radial velocity drift at a maximum of $1/10^{\mathrm{th}}$ of the temporal variability level. This implies that the maximum acceptable, uncalibrated instrumental RV drift is on the level of $\pm 40 \mathrm{m/s}$ over a full transit, to account for all planet orbital velocities. We would like to stress here that we are being conservative in our requirements, as most planets will move much faster and will thus be less affected than our extreme reference case.

\subsection{Instrumental Profile stability}
\begin{figure}
   \centering
   \includegraphics[width=\columnwidth]{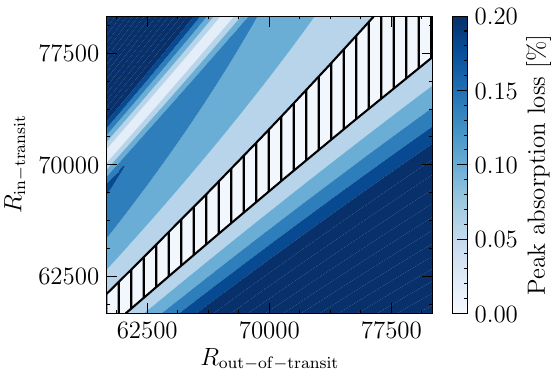}
      \caption{Simulated offset in the peak absorption depth for spectral resolution changes over time. For simplicity, we assumed a fixed spectral resolution for all in-transit spectra and a fixed spectral resolution for all out-of-transit spectra. The shaded area denotes the same .04\% requirement as for radial velocity drifts in Figure~\ref{fig:drift}. The white band of low loss in the upper left region of the Figure is a region that satisfies the requirement but is infeasible. It is introduced by stellar lines being blurred to a degree that it incidentally results in a proper planetary helium line retrieval. As such we discard this region of the parameter space.
              }
         \label{fig:R_drift}
   \end{figure}

Besides radial velocity stability, the instrumental profile (IP) of a spectrograph can also vary over time. For slit-fed spectrographs, the IP or line-spread function (LSF) will strongly vary with the seeing and centering of the target on the slit. We ran simulations for single in- and out-of-transit spectra at different spectral resolutions, which for one wavelength essentially sets the extent of the LSF. For this grid of resolutions, we computed the offset in peak absorption in the same manner as for the radial velocity drifts. The resulting contour can be found in Figure~\ref{fig:R_drift}. 
From Figure~\ref{fig:R_drift} we can conclude that the maximum acceptable change in $R$ is not compatible with a slit-type spectrograph. To have decent efficiency, it is important that a slit has an angular size of at least 0.7" on the sky. At any good site, seeing will likely be close to 1". However, if in excellent conditions the seeing drops to a size smaller than the slit size, the spectral resolution will see a jump -- introducing an offset in the final retrieved helium absorption spectrum of the planet. As such, we can conclude that NIGHT will need to be a fibre-fed spectrograph -- to keep slit sizes, and thus spectral resolutions fairly constant. Besides, we want to design NIGHT in a way that our in-transit and out-of-transit spectra are not shifted by more than 40 m/s (see Section~\ref{sec:wlstab}). This can be achieved by putting (part of) the optics in a pressure and temperature-stable environment and/or calibration techniques. Only relying on a simultaneous wavelength reference and not imposing restrictions on environmental stability could prove difficult as the current design only accounts for two fibres. Since we would like to measure a simultaneous sky reference besides the stellar spectrum, we do not have the option to also continuously measure a calibration source. We could choose an observational strategy where we make a calibration exposure multiple times per night but this would not guarantee the elimination of spurious drifts in-between those calibration moments. A safer option is thermal, and pressure control. The exact precision to which we would need to control the environment will be determined in a later stage after a detailed thermal and pressure change analysis of our mechanical design. However, given the not-so-stringent stability requirement, we expect this to be less than for previously built high-resolution spectrographs.

\subsection{Fibre modal noise}
Multimode fibres can introduce a phenomenon called modal noise, also sometimes referred to as speckle noise. In a multimode fibre, light propagates through various paths and modes. These modes represent different spatial paths the light takes through the optical fibre. Each single mode has its own unique phase and angle of propagation. In multimode fibres, due to their large core diameter, multiple modes can be excited simultaneously. Modal noise occurs due to interferences between these different modes as light propagates. These can be caused by bends, imperfections, and disturbances of the fibre. As a result, for monochromatic light, intensity variations are observed over the output of the fibre, as different fibre modes come together and constructive or destructive interference occurs due to the different phase-offsets of the modes. This effect scales to the power of 2 with wavelength and thus becomes increasingly more problematic for infrared instruments \citep{oliva2019, blind2022}. \citet{blind2022} shows that the number of modes that can be transported through a circular fibre is equal to $V^2/4$ where V scales according to Equation~\ref{eq:modes}:

\begin{equation}
    \label{eq:modes}
    V = 2 \pi \mathrm{NA} \frac{a}{\lambda},
\end{equation}

where NA is the numerical aperture, $a$ the core radius, and $\lambda$ the wavelength.

\citet{frensch2022} show that the NIRPS high-resolution spectrograph \citep{bouchy2017near} exhibits modal noise structures on the level of 1.6\% RMS at 1.55 $\mathrm{\mu}$m with their 29 $\mathrm{\mu}$m few-mode fibre. By the implementation of a fibre agitator, this level could be reduced to just 0.7\%. Modal noise structures can vary over time \citep{oliva2019}, but in the case of NIRPS are stable over the timescale of a few hours \citep{frensch2022}.

In a worst-case scenario, modal noise signatures would fluctuate just enough that flat-fielding does not properly account for them. As a result, modal noise appears as random spectral features. To assess whether modal noise might prohibit our study of time-variable signals in the helium triplet, we use Equation~\ref{eq:modes} to scale the results of \citet{frensch2022} to our wavelength and a fibre core of 60 $\mathrm{\mu}$m.
Using Equation~\ref{eq:modes}, a fiber core size of 60 $\mathrm{\mu}$m, and an NA of 0.125, the total number of modes transmitted by NIGHT at 1.08 $\mathrm{\mu}$m should be equal to $\sim$ 473. \citet{blind2022} state that the NIRPS fibre carries 10 to 35 modes. Given NIRPS's wavelength coverage of 0.95 to 1.8 $\mathrm{\mu}$m, we can assume NIRPS transmits $\sim$ 19 modes at 1.55 $\mathrm{\mu}$m as the number of modes decreases with increasing wavelength. This implies that the NIGHT fibres carry around 25 times more modes. 
\citet{goodman1981} show that modal noise is inversely proportional to the squared root of the number of modes and the signal-to-noise scales linearly with it. As such, modal noise in NIGHT is expected to be a factor of $\sqrt{25} = 5$ times smaller than for NIRPS. This would imply a modal noise RMS of about 0.32\% excluding a fibre agitator. This is just below the temporal spectral depth variations we will attempt to resolve. However, there are three scenarios that could further reduce the magnitude of modal noise:
\begin{itemize}
    \item Natural seeing will help to spatially fill the modes at the telescope focus, reducing modal noise over an exposure.
    \item The modal noise could vary quickly and will even out over a transit when spectra are added.
    \item The modal noise could be stable over a transit duration and when we perform flat-fielding or compute the absorption spectrum, the modal noise divides out (or we can model it if it is stable). 
\end{itemize}

The real modal noise will have to be evaluated when the instrument is put into usage. We would also like to stress that since this is a prototype, we will always have the option of including a fibre agitator (like the one of NIRPS), which would reduce the modal noise.

\section{Preliminary target selection}

   \begin{figure*}
   \centering
   \includegraphics[height=9.5cm]{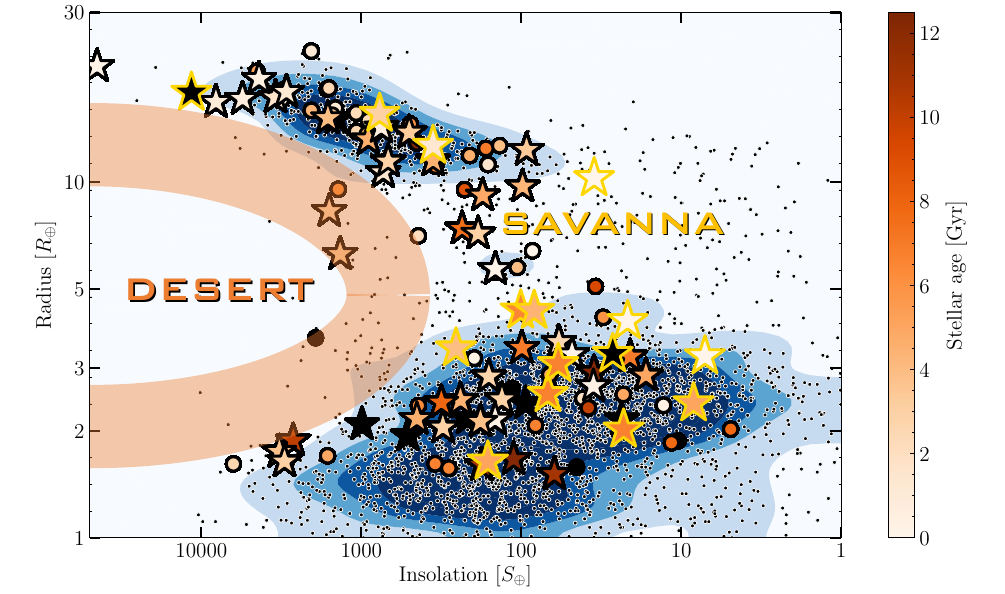}
      \caption{An insolation versus radius plot for known exoplanets. The contours and black dots show the (density) distribution of all known exoplanets to date. The hot Neptune Desert clearly jumps out. We also highlighted the Savanna \citep{bourrier2023}. All 118 NIGHT targets are plotted. Star-shaped markers denote time-variable targets and the gold-bordered markers represent our prime targets. The marker color shows the age of the stellar system in which the planet resides. For systems of unknown age, the marker is colored black.
              }
         \label{fig:night_targets}
   \end{figure*}

Our aim with NIGHT is to observe dozens of transits per year -- requiring a significant fraction of observing time on any telescope. To guarantee this time, a less-competitive 1.5 to 2-meter telescope is preferred over, for example, a VLT-class facility. Besides, we propose NIGHT to be a visitor, fibre-fed instrument. This will allow us to optimise our observing program -- catching opportunities whenever they may occur and maintaining flexibility introduced by the fibre feed mechanism. To determine whether placing NIGHT at a 2-m class achieves our science goals, we built an SNR simulator. This NIGHT simulator allows adjusting telescope parameters like aperture size and atmospheric and telescope efficiency, as well as instrument characteristics – like the slit/coupling efficiency and detector noise levels. For all exoplanets from the NASA Exoplanet Catalog and The Extrasolar Planets Encyclopaedia, we created mock observations. To create synthetic spectra we used a MARCS synthetic photosphere spectral model (4500 K, 4.5 $\mathrm{log} g$) \citep{gustafsson2008grid}, a telluric absorption model \citep{wallacelivingston2003}, and a helium absorption profile model created with the EVaporating Exoplanets Code (EVE) \citep{bourrier20133d} in \citet{allart2019high}.

   \begin{table}[!h]
      \caption[]{SNR simulator input parameters and the resulting number of targets for NIGHT at a 2-meter telescope.}
         \label{tab:simchar}
     $$
         \begin{array}{p{0.5\linewidth}l}
            \hline
            \noalign{\smallskip}
            Parameter      &  Value\\
            \noalign{\smallskip}
            \hline
            \noalign{\smallskip}
            Atmosphere efficiency & 60\% \\
            Telescope optics efficiency & 70\% \\
            Fibre coupling efficiency & 70\%\\
            Instrument efficiency & 60\%\\
            Quantum efficiency & 50\%\\
            Detector read noise & 20\ \mathrm{e^-/pix}\\
            Detector dark current & 1\ \mathrm{e^-/pix/s}\\
            \noalign{\smallskip}
            \hline
            \noalign{\smallskip}
            Targets      &  Number\\
            \noalign{\smallskip}
            \hline
            \noalign{\smallskip}
            Prime targets & 16\\
            Time-variable targets & 66\\
            Total targets & 118\\
            \noalign{\smallskip}
            \hline
         \end{array}
     $$ 
   \end{table}

Building on the pretense of a 2-meter telescope aperture and conservative efficiency values and average detector noise characteristics as listed in Table~\ref{tab:simchar}, we used stellar J-band magnitude, transit duration, and stellar/planet radii – to determine what SNR was achievable for known transiting planets. Consecutively, we put a cut on the SNR to determine for which planets we would be able to observe temporal variations of 0.4\% in absorption depth at 3 sigma between two independent transits as per our science requirement (see Section~\ref{sec:reqs}). This SNR value is $\approx$1073 per pixel, stacked over a full transit. As of now, there are very few significant detections of strong temporal variations in the absorption signatures of helium atmospheres, but given the strong dependence on stellar XUV flux, the stellar wind, and more generally the stellar environment, it is suspected that variations can be large over short timescales, especially for younger systems \citep{lammer2009, lecavelier2012, vidotto2015, vidotto2016, kubyshkina2018, vidotto2020, poppenhaeger2022}.\\

Table~\ref{tab:simchar} lists the resulting number of targets generated by our simulations on the full sky. Targets were pre-selected to have an orbital period <30 days and to have a radius >1.5$\mathrm{R_{\oplus}}$, to filter out hot and strongly irradiated, likely evaporating, gas-rich planets. No stellar spectral type or stellar activity is taken into account for now. The targets are split into 16 prime targets (with a confirmed or theorised exosphere), 66 time-variable targets (which meet the temporal requirement of 0.4\% absorption depth change at 3 sigma significance between two independent transits), and 118 targets total, for which we would be able to detect an absorption signature of 1\% at 5 sigma significance (SNR $\approx$708) in a single transit. Note that the time-variable targets are a subset of the total amount of targets, and the prime targets are a subset of the time-variable targets. More detections could be reached for low-SNR targets by stacking observations from multiple transits, which could be interesting for targets that transit frequently. Stacking multiple transits will result in an improved signal, but potentially limit temporal monitoring, depending on the exact timescales involved with the short-term fluctuations of these atmospheres. We would like to stress that absorption depths much deeper than 1\% are not unusual and have been detected many times \citep{allart2018spectrally, allart2019high, zhang2023_32b, bello-arufe2023}, and as such, our estimates are conservative. Later on in this chapter, we will look at how the telescope's location influences the number of observable targets.

Figure~\ref{fig:night_targets} shows the insolation versus radius distribution of our targets. Insolation is a measure of the amount of irradiation a planet receives at the top of its atmosphere from its host star. This depends on the separation between the star and the planet and the stellar type. For hydrodynamical expansion and escape induced by stellar radiation, we know that XUV flux is the main driver. We use bolometric irradiation as a first-order proxy as the XUV irradiation is unknown for most planets. Our targets cover well the edges of the hot Neptune Desert, covering a large range of insolations from 5 to 2$\cdot 10^4$ $\mathrm{S_\oplus}$ in the sparsely populated area of the Neptune-size range. This region is especially interesting as we expect Neptune-size planets to be most sensitive to atmospheric escape.

\subsection{Telescope choice}

\begin{figure}
   \centering
   \includegraphics[height=5.5cm]{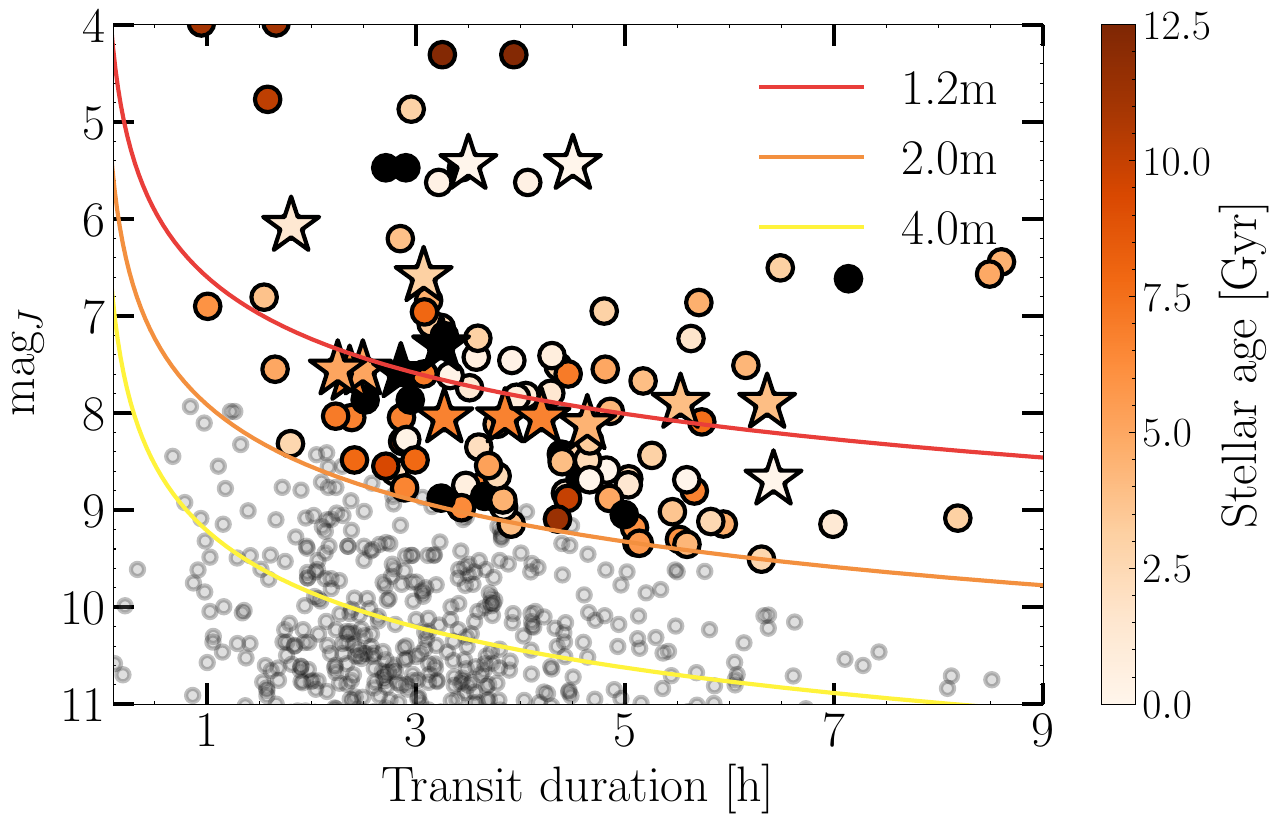}
      \caption{Limiting magnitudes for a 5 sigma detection of a 1\% excess absorption feature in the helium I triplet versus transit duration. Limits are plotted for three different telescope aperture diameters. Small, faded, circular markers show all known transiting planets with a period $<$30 days and radius $>$1.5$\mathrm{R_\oplus}$. Circular markers show our 118 NIGHT targets and the star-shaped markers denote our prime targets (16 total). As in Figure~\ref{fig:night_targets}, the color of the marker shows the age of the stellar system.
              }
         \label{fig:lim_mag}
   \end{figure}

Since NIGHT will be a visitor instrument, we also analysed what potential gains could be made in terms of the number of observable targets by moving the instrument to a larger telescope. In Figure~\ref{fig:lim_mag} we can find the limiting magnitudes for 3 telescope apertures of 1.2, 2.0, and 4.0 meters in diameter, with the same 5 sigma requirement as before. The limiting magnitude depends on the duration of the transit, as for longer transits more exposures of equal integration time can be taken, boosting signal-to-noise by stacking the spectra. For a larger statistical sample, a telescope aperture of 2 meters or larger is preferred. Some faded dots lie within the observable magnitudes but are not part of the full target list. To determine the 5 sigma threshold, we only take into account the duration when the planet is in full transit (after ingress and before egress). As such, depending on the planet and stellar radius, some targets close to the limiting magnitude did not meet our requirement. 

It is important to realise that in this scenario we assume similar fibre coupling efficiencies for all telescope apertures. This is an unrealistic scenario as the etendue increases for larger telescopes and as such, a fixed fibre size will have lower coupling efficiencies on larger telescope apertures. We can make a comparison based on the plate scale, which relates the angular separation on the sky to the physical size in the focal plane of the telescope, given by Equation~\ref{eq:pl_sc}:
\begin{equation}
    \label{eq:pl_sc}
    p \sim \frac{206265}{f},
\end{equation}

where $f$ is given in mm and $p$ in "/mm. We see that the plate scale is inversely proportional to the focal length of the telescope. Now let us define $s$, the size of 1" in $\mathrm{\mu m}$ on the focal plane. We can transform Equation~\ref{eq:pl_sc} into Equation~\ref{eq:s}:
\begin{equation}
    \label{eq:s}
    s = \frac{D}{206265\cdot \mathrm{F}}\cdot 10^3,
\end{equation}
where $D$ is the diameter of the telescope in mm and F is the inverse of the f-number of the telescope. We can subsequently apply a conversion factor of F-ratios from telescope to fibre:
\begin{equation}
    \label{eq:ss}
    s = \frac{D\cdot 10^3}{206265\cdot \mathrm{F}_{tel}}\cdot\frac{\mathrm{F}_{fib}}{\mathrm{F}_{tel}}
\end{equation}

Assuming a constant F-ratio of the telescope and fibre, we find that the size of a 1-arcsecond star (in a seeing-limited scenario) in the focal plane is proportional to the diameter of the telescope. This value is directly related to the coupling efficiency into the optical fibre as the size of the fibre preferably allows for most of the starlight to pass through. For larger aperture telescopes, we could sustain a similar level of coupling efficiency by demanding a faster F-ratio. Another possibility we considered was slicing the pupil of the input fibre, but given this will significantly affect the instrumental profile (IP), we dismissed it as an option. Overall, we conclude that it is preferred to optimise NIGHT for the largest fibre size we can accommodate while keeping the instrument reasonably compact to meet our requirements and accordingly choose our telescopes wisely. In this decision, we will need to make the trade-off between the expected coupling efficiency, derived from the telescope parameters given our spectrograph, and target brightness and visibility.

\subsection{Observing program}

\begin{figure}
    \centering
    \includegraphics[width=\columnwidth]{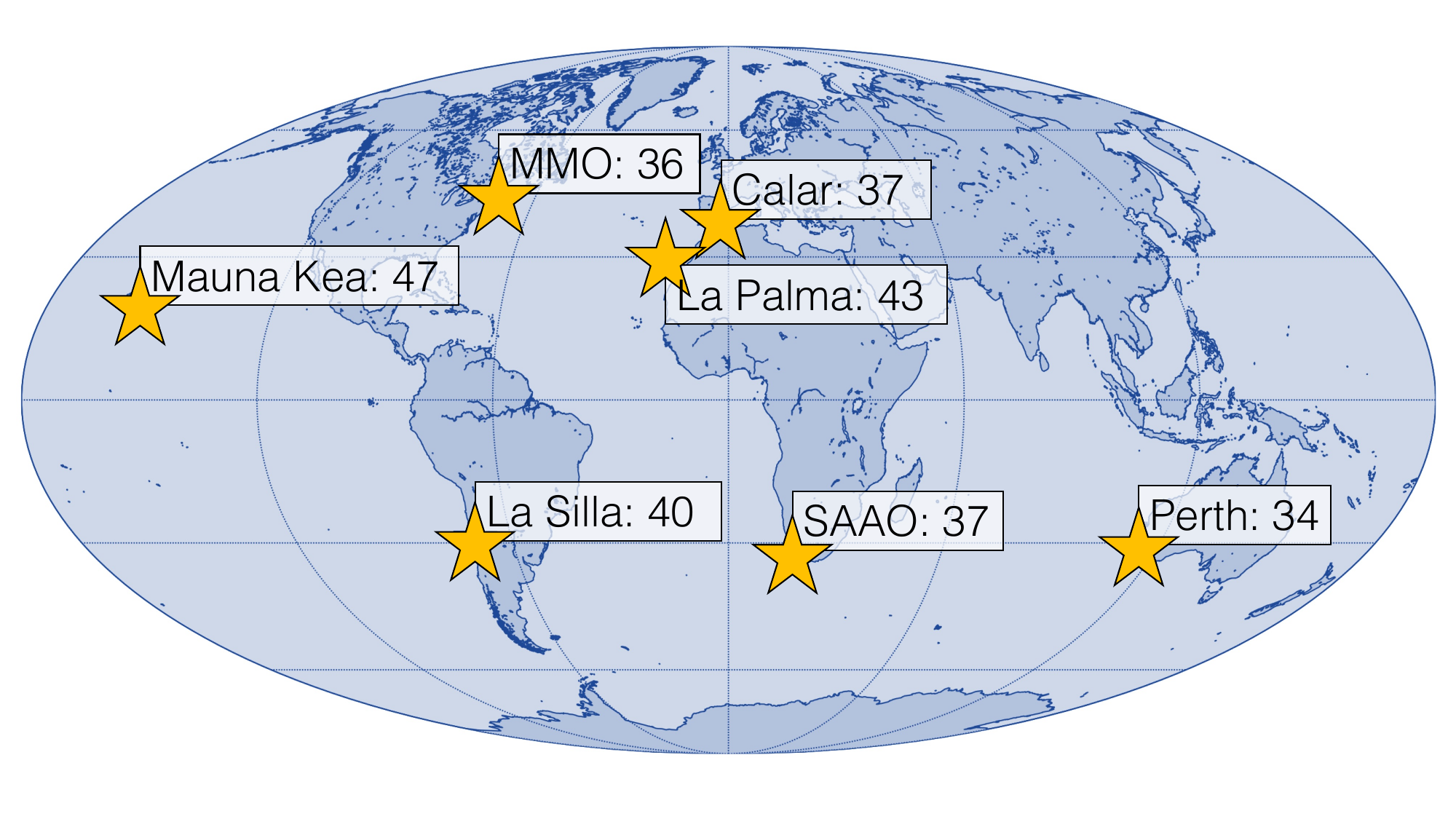}
        \caption{A world map showing for how many exoplanets of our time-variable target list (total: 66 planets) we can observe at least 2 transits in the year 2024 from various sites. To acquire these numbers, we made a conservative estimate: only transits that are fully visible in astronomical twilight from ingress to egress at airmass <2 were taken into account.
              }
        \label{fig:night_map}
    \end{figure}

We performed a target transit visibility analysis of all temporal and prime targets for the year 2024 for various sites on Earth and found no clear preference for any specific location to conduct our observations. The results of this analysis for our time-variable targets can be found in Figure~\ref{fig:night_map}. In general, we conclude that transits are rare events for the class of planets we target. Most of our targets transit only a few times a year, and for half of our targets only once. This is mostly induced by requiring a good alignment with Earth nighttime to catch a full transit (which was taken into account in our simulations). Note that in these simulations we only consider the time of transit, and ignore the required baseline. In case of a long transit that takes up the entire night, we could always opt for baseline measurements the night before and/or after the transit. \citet{mounzer2022} show that baseline measurements on consecutive nights still allow for very accurate retrievals, even in the presence of tellurics.

\begin{figure}
   \centering
   \includegraphics[width=\columnwidth]{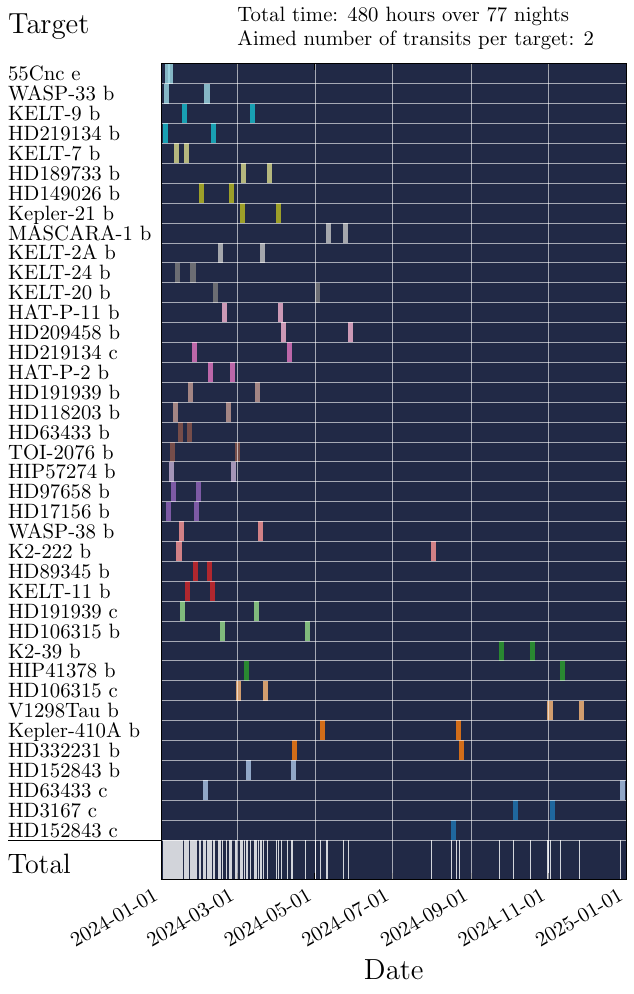}
      \caption{A preliminary observing schedule for all targets listed in Appendix~\ref{Appendix1} for MMO. Some targets of the full table are not included because they are i) not observable from this latitude, or ii) no two full transits are observable from this location in 2024. Our automatic scheduler code computes all observable transits for a given list of targets and location on Earth. After, it filters for full transits < airmass 2. The transits are automatically scheduled such that no transits overlap and priority is given to targets with the least amount of full transits. The total observation time required includes an out-of-transit baseline of 50\% the length of the transit duration.
              }
         \label{fig:obs_prog}
   \end{figure}

We expect that within one year of surveying, we can build up a statistical, temporal sample of extended helium atmospheres for 30 to 40 planets, with at least 2 transits per target. We would need to be allocated about 70 nights per year for this survey to acquire 2 transits per target. An example of a possible observing program can be found in Figure~\ref{fig:obs_prog}. This observing program is based on the assumption that NIGHT will be placed at Mont Mégantic Observatory (MMO), a 1.6-meter F/10 telescope. Depending on the total time allocated to NIGHT, this number could increase to over 50 planets for one site in one year, with temporal data depending on the achieved SNR for each transit and the strength of any temporal variations. The total number of observed atmospheres can be even further increased by stacking multiple transits. Furthermore, if the science case is expanded to monitoring stellar helium absorption, NIGHT observations could easily fill up all available time on a telescope. Monitoring stellar helium absorption is likely useful to better constrain the shape of the lines in out-of-transit spectra, and with that, more accurately retrieve the planetary absorption features. Potential temporal variations of the depth and shape of these lines in the star itself can be better understood and possibly predicted if monitored over longer baselines than just the transit. As such, we aim for part of our strategy to consider long-term monitoring of stellar He I absorption.

\section{Instrument description}

In Table~\ref{tab:reqs} all technical requirements are summarised. They are a result from all simulations and trade-offs presented in this work.

   \begin{table}
      \caption[]{Summarised requirements.}
         \label{tab:reqs}
     $$ 
         \begin{array}{p{0.5\linewidth}l}
            \hline
            \noalign{\smallskip}
            Requirement      &  \mathrm{Value}\\
            \noalign{\smallskip}
            \hline
            \noalign{\smallskip}
            Wavelength range & 10810-10850 \AA\\
            Spectral resolution (R) & 70,000-75,000\\
            Sampling (pixels/FWHM) & \mathrm{at\ least}\ 2\\
            Wavelength calibration & \mathrm{Ur-Ne\ \&\ tellurics}\\
            Wavelength precision & <40\ \mathrm{m/s}\\
            Nightly RV drift & <40\ \mathrm{m/s}\\
            Slit type & \mathrm{optical\ fibre}\\
            Number of slits & 2\ \mathrm{(one\ science,\ one\ sky)}\\
            Instrument frontend efficiency & >70\%\\
            Instrument backend efficiency & >60\%\\
            Telescope diameter & 2 \mathrm{m}\\
            Quantum efficiency & >50\%\\
            Read noise & <20\ \mathrm{e^-/pix}\\
            Dark current & <1\ \mathrm{e^-/pix/s}\\
            \noalign{\smallskip}
            \hline
         \end{array}
     $$ 
   \end{table}

\subsection{Spectrograph type}
\label{sec:spectro}
For NIGHT we considered various spectrograph types: \\ A Fourier-transform spectrograph (FTS), a Virtual Image Phased Array (VIPA), and a grating spectrograph. An FTS was quickly discarded due to the intrinsic multiplexing deficit, boosting our photon noise. The VIPA has been proposed to be used in exoplanetary science HR spectrographs by \citet{carlotti2022sky}, but we dismissed it as an option due to expected worsened radial velocity stability. The Fabry-Perot-like dispersion has strong thermal and pressure dependence, and besides, it has an intrinsically small free spectral range (FSR). As such, a grating-based spectrograph was decided on.

   \begin{figure*}
   \centering
   \includegraphics[width=0.48\textwidth]{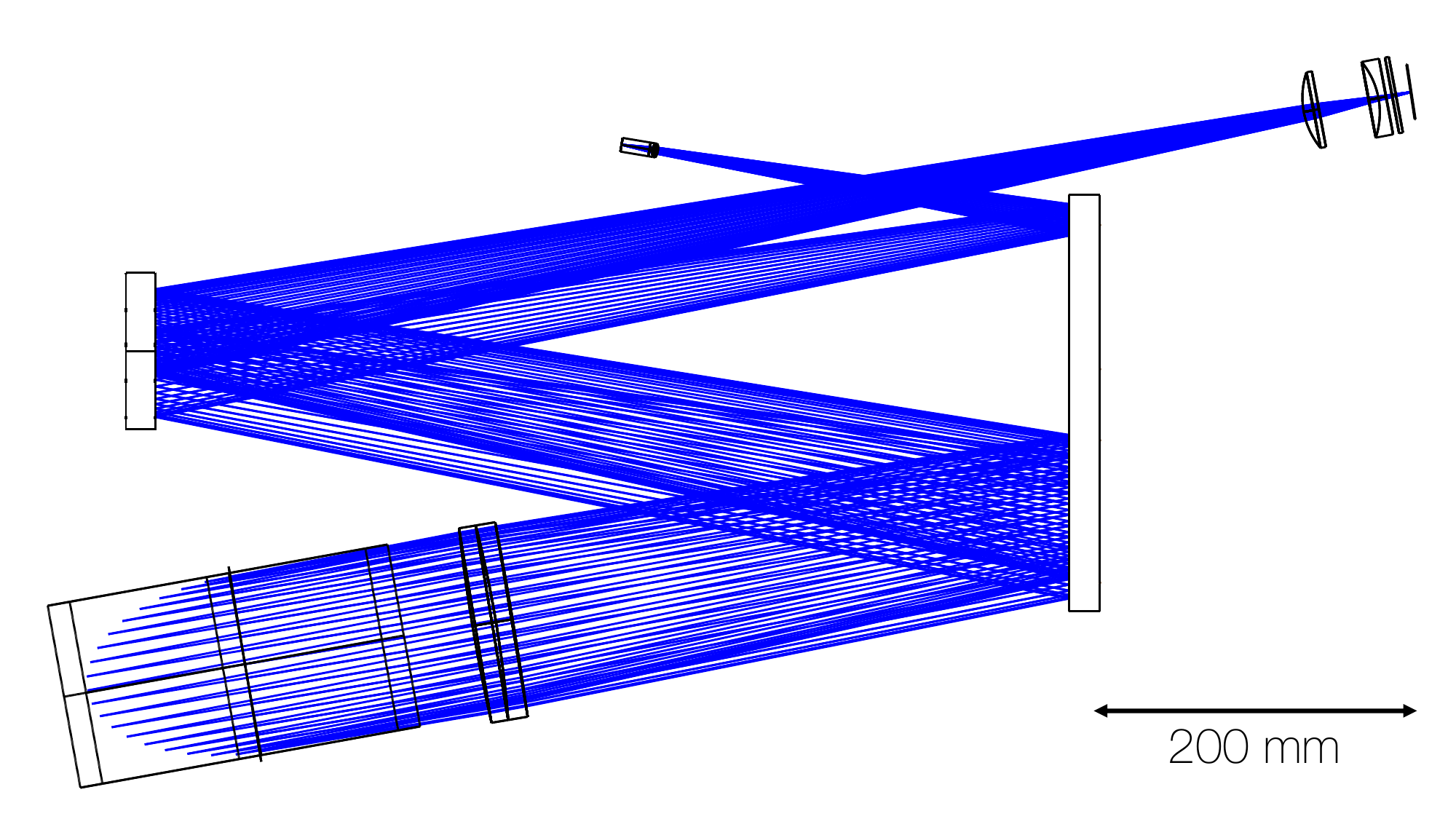}
   \includegraphics[width=0.48\textwidth]{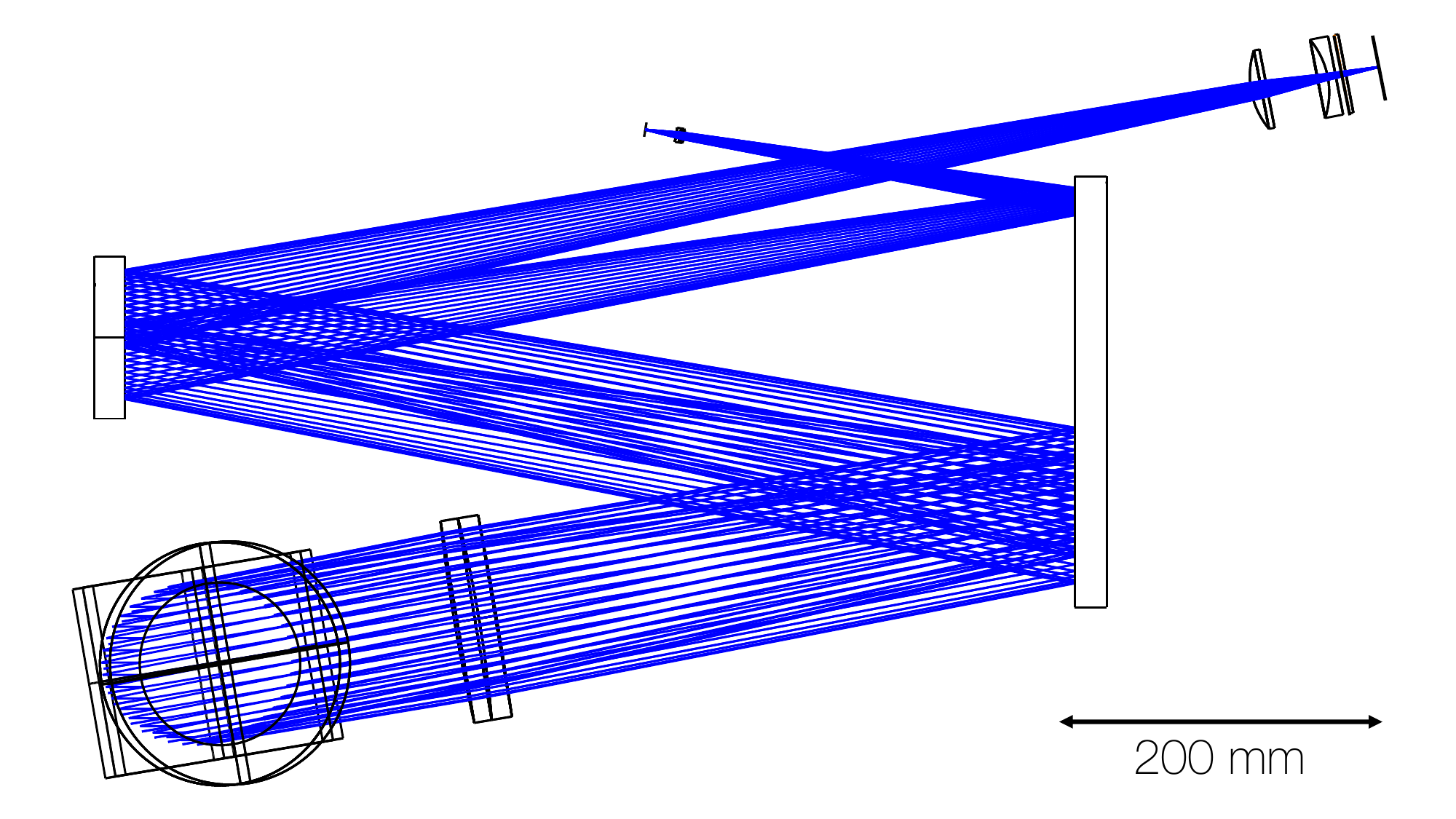}
   \caption{Optical design of NIGHT with an Echelle grating in a). In b) we find the optical design with a VPH grating in double pass.
              }
         \label{fig:opt}
   \end{figure*}

It has always been our aim to design NIGHT as a compact, cost-efficient instrument. To keep costs at an acceptable level, we decided to use mostly off-the-shelf optical elements. While this keeps costs low, it also introduces less flexibility in the design. Having set our requirements on spectral resolution and telescope size, our trade-offs for the optical design progressed in the following way:
\begin{enumerate}
    \item set slit and collimator size based on requirements;
    \item find an off-the-shelf collimator;
    \item find a suitable grating with the right dispersive power;
    \item find remaining optical components for the fibre injection and camera.
\end{enumerate}
Currently, two designs for NIGHT exist -- they are overall very similar but contain different gratings. One design utilises an R2 Echelle grating in 52nd order with 31.6 lines/mm, put 1-degree off-Littrow to center the blaze on our passband. The second consists of a VPH grating in first order with 1406 lines/mm used in double-pass by the addition of a flat mirror. All other optical elements are the same in both designs.

Any grating-type spectrograph's size is guided by the slit size, dispersive power, and collimator focal length, ignoring (opto)mechanical structures. Namely, given a certain dispersive power and slit size, this fixes the size of the collimated beam, and with it the size of the grating.

While it may seem that there are quite a few free parameters, most are either fixed or related for a given spectral resolution, and/or grating type. For example, for an echelle type grating in Littrow configuration, we can relate the maximum achievable spectral resolution $R_{\mathrm{max}}$ to the telescope diameter ($D_\mathrm{Tel}$), the angle-of-incidence on the grating ($\beta$), the height of the grating ($h$), and slit size ($\phi$) in arcseconds following Equation~\ref{eq:spectro3} \citep{pepe2000}:

\begin{equation}
    \centering
    R_{\mathrm{max}} = \frac{2 \cdot \mathrm{tan}\beta \cdot h}{\phi \cdot D_{\mathrm{Tel}}}.
    \label{eq:spectro3}
\end{equation}

The angle $\beta$, together with the total amount of illuminated grooves, and the spectral order observed determines the dispersion. We can quickly see that two of the values in Equation~\ref{eq:spectro3} will be more-or-less fixed, independent of how we design our spectrograph. These are the slit size and telescope diameter. For example, for 1.2 arcsecond on the sky, a telescope diameter of 200 cm, a spectral resolution of 75,000, and a tan$\beta$ value of 2, our grating will have a height of $\approx$ 22.3 cm. With the given angle of incidence on the grating (tan$\beta$ is 2), this gives a collimated beam height of $\approx$ 10.1 cm.

Replacing the Echelle grating with a different grating type working in first order with a (much) higher groove density, like a volume-phase-holographic (VPH) grating, is also an option as long as Equation~\ref{eq:spectro4} satisfies the threshold on resolving power -- the order $m$, times the number of illuminated lines $N$:

\begin{equation}
    \centering
    R = m \cdot N,
    \label{eq:spectro4}
\end{equation}
   
since we would like to reach the same spectral resolution. 
For both of our designs, the collimated beam diameter incident on the grating is similar as pushing the groove density of the VPH to higher values would result in too high losses. Please see Figure~\ref{fig:opt} for reference. Although the design is currently undergoing a design review (DR), further testing in our optical lab of an off-the-shelf VPH manufactured by Wasatch Photonics will inform us if a double-pass VPH performs up to expectations. To our knowledge, this grating type has never been used in double-pass to achieve a spectral resolution of $R = 75,000$. The manufacturer is able to produce the custom VPH grating for NIGHT with our specifications. Echelle-type spectrographs regularly reach these resolutions but are frequently cross-dispersed due to the overlap of spectral orders. In first order, this is not a significant problem because the FSR is much larger. Namely, the FSR is given by:

\begin{equation}
    FSR = \frac{\lambda}{m}.
\end{equation}

\begin{figure}
   \centering
   \includegraphics[width=\columnwidth]{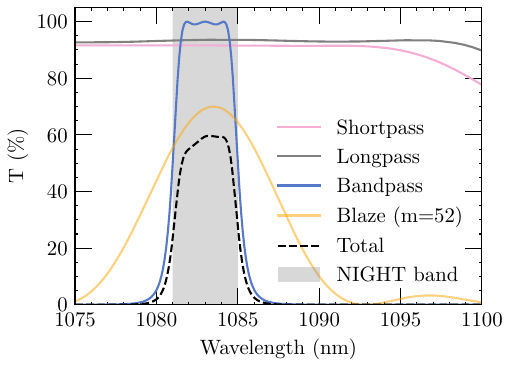}
      \caption{Total transmission of the custom filters and echelle grating blaze. We can see that the peak transmission reaches close to 60\%, whereas it drops off to a minimum of about 20\% at the sides of the band.
              }
         \label{fig:filters}
   \end{figure}

\begin{figure}
   \centering
   \includegraphics[width=\columnwidth]{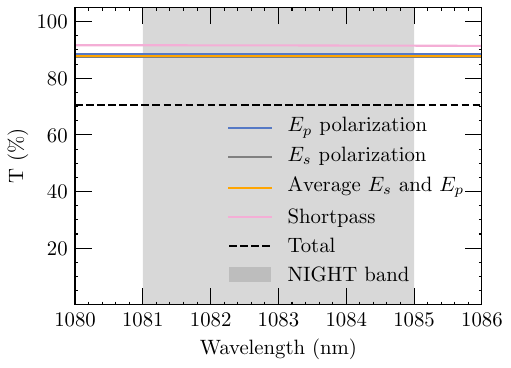}
      \caption{Total transmission of the custom VPH Dickson grating designed by Wasatch Photonics in double pass (ignoring losses at the flat mirror). Because no custom bandpass filters are required, and the grating has a higher diffraction efficiency than a typical Echelle grating, the total transmission over the entire NIGHT passband is higher than for the Echelle layout at an average of about 71\%. Note that we use a different wavelength scale than in Figure~\ref{fig:filters}.
              }
         \label{fig:vpheff}
   \end{figure}

For the Echelle grating in 52nd order, this implies an FSR of about 21.2nm, centered on our wavelength band. This implies that many higher-, and lower orders will be overlapping, requiring additional (custom) filters in the front-end of NIGHT -- increasing cost and reducing efficiency. A single narrow bandpass filter would not be sufficiently blocking all wavelengths for which the detector is sensitive, as to why we require 3 separate filters. The accumulation of three custom filters and the expected Echelle diffraction efficiency can be found in Figure~\ref{fig:filters}. It consists of a long pass filter to block lower orders, a short pass filter to block higher orders and heat radiation, and a narrow bandpass filter to block the surrounding orders. For comparison, the expected efficiency of the VPH grating in double pass can be found in Figure~\ref{fig:vpheff}. Because no additional narrow bandpass and long pass filter are required, the higher peak diffraction efficiency (compared to an Echelle grating), and the much broader FSR, the overall transmission is higher. In terms of transmission, a VPH in double-pass is preferred for NIGHT, but further testing is required to establish whether no residual optical effects are introduced by this setup. Independent of the grating, both designs will work, albeit with a lower transmission than set by our requirements in the case of the Echelle solution. 


\subsection{Optical Performance}
\begin{figure}
   \centering
   \includegraphics[width=0.95\columnwidth]{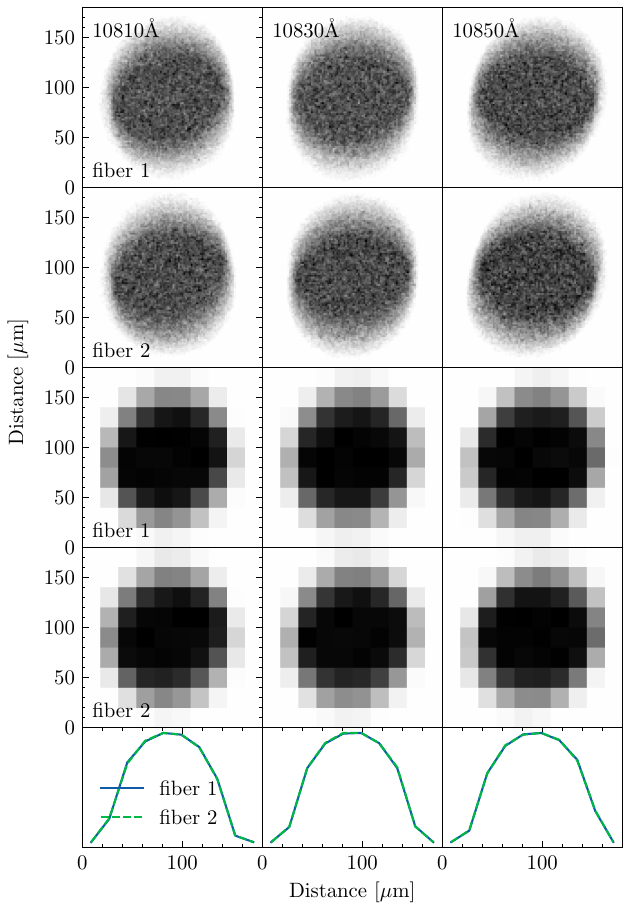}
      \caption{The PSF at high and low resolution at three different wavelengths for both fibres in the echelle design. This is the PSF at the image/detector plane of the instrument. The low-resolution PSF is sampled at the pixel pitch of the proposed detector for NIGHT. In the lower panels, we show the 2D extracted PSF, stacked over the pixel rows. Note that for both fibres, the 2D PSF is very similar in shape and the lines overlap.
              }
         \label{fig:PSF}
   \end{figure}

Both optical designs of NIGHT were optimised with Zemax OpticStudio. All lenses and fold mirrors were chosen as off-the-shelf components from Edmund Optics. The bottleneck in our design was the collimator. From Section~\ref{sec:spectro} we recall that the size of the collimated beam would need to be $\approx$10.1cm. Off-the-shelf, off-axis parabolic mirrors typically only go to a 10cm diameter for a focus of F/8 or slower, which would just not accommodate our beam size. It is important to have an F/8 beam or slower for the collimator to minimise optical aberrations. In terms of lenses, only very few off-the-shelf doublets exist that accommodate our beam size. The only lens that we found, that did not introduce too many optical aberrations, was a F=1900.2mm doublet from Edmund Optics. The long focal length of this collimator has the advantage that it reduces optical aberrations but has the disadvantage that it increases the total size of the instrument. Through the placement of 3 fold mirrors, of which 2 in double-pass, we were able to keep the size of the instrument fairly compact. To quantify the expected optical performance, MCMC tolerance analysis was run on both optical designs, both for manufacturing and assembly tolerances. The merit function that we defined in Zemax OpticStudio computed the total enclosed energy in a pixel of 18$\mathrm{\mu m}$ in the dispersion direction, with some constraints on the placement of the optics, for two 60$\mathrm{\mu m}$ 0.125NA circular fibres as entrance slits (one science fibre, one sky fibre). The fibre type was chosen as it is the largest fibre that will allow us to meet our spectral resolution requirement with the given beam constraint. We ran a least-squares optimisation algorithm to maximise the total enclosed energy -- pushing the spectral resolution while allowing the spectrum to spread over more rows of pixels. The point-spread functions (PSFs) of the optimised echelle design can be found in Figure~\ref{fig:PSF}.

   \begin{figure*}
   \centering
   \includegraphics[width=0.92\columnwidth]{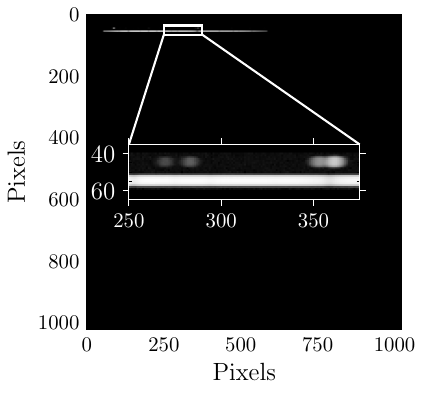}
   \includegraphics[width=0.9\columnwidth]{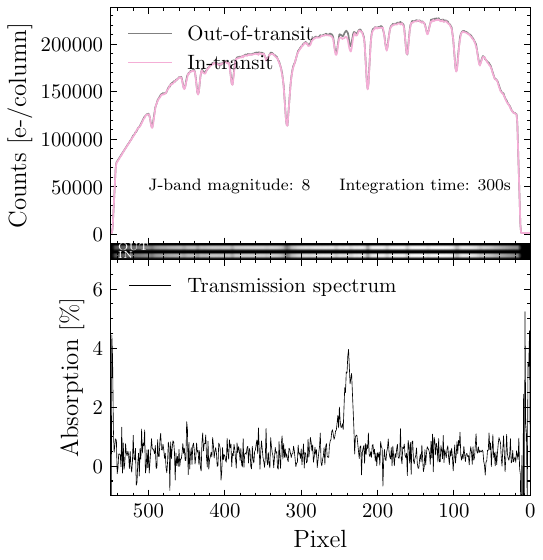}
      \caption{On the left-hand side we show the full-frame ray trace from both fibre feeds of the Echelle spectrograph on a 1K x 1K detector. The upper spectrum shows the stellar spectrum and the lower a Ur-Ne calibration spectrum. The zoom-in shows the intensity in logarithmic scaling to improve the visibility of the Uranium lines in the calibration spectrum. On the right-hand side, we show the extracted spectra from two simulated exposures -- one out-of-transit and one in-transit. The synthetic spectra consist of a MARCS photospheric model of a 4500K, $\mathrm{log}g = 4.5$ star, a telluric absorption spectrum, and a re-scaled WASP-107\;b helium signature. The WASP-107\;b signature was re-scaled to 3\% peak excess absorption. We assume a J-band magnitude of 8, an integration time of 300 seconds, and a telescope size of 2 meters. The telescope and instrument efficiencies/characteristics and detector properties are taken from Table~\ref{tab:simchar}, Figure~\ref{fig:filters} \& Figure~\ref{fig:PSF}.
              }
         \label{fig:HAWAIIdet}
   \end{figure*}

From the PSFs, we find that we are at near-diffraction-limited performance and reach a spectral resolution of $R\sim70,000$. PSFs look slightly elongated in the non-dispersion direction since we did not optimise for enclosed energy in this direction. From the extracted PSFs we built a Python-based instrument simulator. The resulting detector image can be found in Figure~\ref{fig:HAWAIIdet}. 

From this simulated detector image and synthetic spectra, we extracted spectra by stacking the illuminated rows. The resulting extracted spectra can be found in Figure~\ref{fig:HAWAIIdet}. We can see that for a J-band magnitude of 8, in a 5-minute exposure, we can easily extract an absorption signature of 3\%. Of course, typically many exposures are stacked, increasing the signal-to-noise of the spectra -- allowing us to meet our science requirements and survey the planets as highlighted in Figure~\ref{fig:night_targets}. The instrument simulator confirms that with the current optical design, we meet our requirements on spectral resolution and sampling with values of $R=70,000$ and 2.9 pixels/FWHM, on the bandwidth (10,810 to 10,850\AA), and are in the range of SNRs we expect.

\subsection{Detector}
Above 1 micron in wavelength, few detectors with a decent quantum efficiency (Q.E.), and low dark noise exist. Silicon-based CCD detectors become transparent above 1 micron, and even with a thick substrate, 1.08 microns is not feasible with a Q.E. of a few percent. Experiments with Indium-Gallium-Arsenide (InGaAs) detectors have shown good Q.E., but too high dark current for high-resolution spectroscopy in an astronomical context \citep{nelson2006development, schindler2014characterization, sullivan2014near}. Cryogenically cooled Mercury-Cadmium-Telluride (HgCdTe) detectors supply good Q.E. (>50\%) and low dark noise (<1 e-/p/s) and are widely used in the field of near-infrared spectroscopy \citep{finger2008performance}. Well-known detector arrays are for example the HAWAII-RG series manufactured by Teledyne, coming in 1kx1k, 2kx2k, and 4kxk4k formats. Older models include the discontinued HAWAII series manufactured by Rockwell. Newly built, science-grade detectors fall outside of our current budget. However, given the narrow wavelength range, only a small (part of a) detector array is needed of about 650x25 pixels (see Fig.~\ref{fig:HAWAIIdet}). As such, some engineering grade C detectors suffice our requirements, depending on the exact deficits. The fact that we only require a small detector footprint of a high-performance IR detector is what keeps the costs of NIGHT affordable. Two engineering-grade detectors are currently being tested for cosmetics at the Department of Physics at the University of Montreal. We aim to cryogenically cool the detector with an Infrared Laboratories Dewar, either liquid nitrogen or Stirling engine based.

\subsection{Timeline}
The NIGHT VPH grating and HAWAII detector are currently being tested at the Department of Astronomy at the University of Geneva and the Department of Physics at the University of Montreal. After these tests and a final design review, we are planning to acquire all components to build the spectrograph this year (2023). We are aiming for first light during the second half of 2024.

\section{Conclusions}
We have researched the feasibility of realising a compact, cost-efficient, high-resolution spectrograph named NIGHT, optimised to spectrally and temporally resolve excess absorption in the He I triplet at 1083nm during exoplanet transits. 

By using past detections of a helium signature at high spectral resolution and the currently known exoplanet population, we set technical requirements for the instrument and designed a spectrograph from mostly off-the-shelf components. NIGHT is a stabilised, fibre-fed spectrograph at a spectral resolution of $R\sim70,000$. The spectral range covers the 10,810--10,850 $\AA$ band and it will be wavelength-calibrated using Ur-Ne and telluric lines to reach 40 m/s precision. As of now, the grating type for NIGHT remains under study with 2 possibilities: an echelle type grating with custom order sorting filters or a VPH grating in double-pass. However, we have set the other optical components and these are readily available off-the-shelf. Detector testing is currently underway at the Department of Physics at the University of Montreal and we hope to have all components at the Observatory of Geneva before the end of 2023. The first light is aimed for the year 2024. 

Based on our simulations for target selection, we estimate that with NIGHT on a 2-meter class telescope, we could survey and temporally monitor over 100 planets. With the current design of NIGHT, we should be able to detect an excess absorption signature of 1\% for 118 targets in a single transit at 5-sigma significance. This number would naturally increase with newly discovered planets and by stacking multiple transits. With the same aperture size, NIGHT could observe potential temporal variations in the shape and extent of these atmospheres at the level of 0.4\% in absorption depth at 3-sigma significance in between 2 transits for 66 planets. Besides the dedicated NIGHT survey, its flexibility as a visitor instrument on a small telescope makes it an ideal follow-up instrument for helium detections made with the Hubble Space Telescope and JWST, or even simultaneous observations. NIGHT also introduces the ability to monitor targets for extended periods of time after an initial detection with more competitive ground-based HR spectrograph, covering wider science cases, like NIRPS. NIGHT will thus allow increasing the statistical sample of extended atmospheres around exoplanets and monitoring their temporal variability. 

\section*{Acknowledgements}
This work has been carried out under the Swiss National Science Foundation (SNSF) grant nr. 184618. This work has partially been carried out within the framework of the NCCR PlanetS supported by the SNSF under grants 51NF40$_{}$182901 and 51NF40$_{}$205606. Besides, this project has received funding from the European Research Council (ERC) under the European Union's Horizon 2020 research and innovation programme (project {\sc Spice Dune}, grant agreement No 947634). This work was supported in part through a grant from the Fonds de Recherche du Qu\'{e}bec - Nature et Technologies (FRQNT). This work was also partially funded by the Institut Trottier de Recherche sur les Exoplan\`{e}tes (iREx). R. A. is a Trottier Postdoctoral Fellow and acknowledges support from the Trottier Family Foundation. The authors would like to thank Omar Attia for supplying his Python code that allows for the easy extraction of data from the exoplanet archives. The authors would also like to explicitly thank Christian Schwab and other reviewers for their input throughout the conception of NIGHT. Lastly, we would like to thank the anonymous referee for their thorough comments. 
This is a pre-copyedited, author-produced PDF of an article accepted for publication in the Monthly Notices of the Royal Astronomical Society published by Oxford University Press on behalf of the Royal Astronomical Society following peer review. 

\section*{Data Availability}
The data underlying this article will be shared on reasonable request to the corresponding author.



\bibliographystyle{mnras}
\bibliography{manuscript} 




\appendix

\section{Time-variable targets for NIGHT}
\label{Appendix1}

\begin{table*}[]
    \begin{minipage}{\textwidth} 
	\begin{center}
            \scriptsize
		\begin{tabular}{c|c|c|c|c|c|c|c|c|c|c}
                Name\footnote{Prime targets in bold.} & Ra [deg] & Dec [deg] & $\textrm{mag}_J$ & SNR\footnote{Note that this is the achievable SNR per pixel based on the total collected flux at the detector. It is achieved by stacking all frames through an entire transit. It also includes detector noise.} & period [d] & r [$r_{\oplus}]$ & I [$S_{\oplus}$]\footnote{Note that for some planets we do not know the insolation. This is a result of not knowing the stellar bolometric luminosity and/or stellar radius.} & Spec. type & $T_{1-4}$[h] & Reference \\
                \hline
			HD 1397 b & 4.45 & -66.36 & 6.4 & 3648 & 11.54 & 11.5 & 356.0 & G5 III/IV & 8.6 & \citet{2019AA...623A.100N} \\
			HD 3167 c & 8.740 & 4.380 & 7.5 & 1665 & 29.840 & 2.9 & 17 & K0 V & 4.810 & \citet{2016ApJ...829L...9V} \\
			HD 5278 b & 12.550 & -83.740 & 6.9 & 2219 & 14.340 & 2.4 & 132 & F V & 4.800 & \citet{2021AA...648A..75S} \\
			K2-222 b & 16.460 & 11.750 & 8.4 & 1073 & 15.390 & 2.4 & 94 & G0 & 4.390 & \citet{2018AJ....155..136M} \\
			\textbf{WASP-33 b} & 36.710 & 37.550 & 7.6 & 1146 & 1.220 & 17.9 & 11540 & A5 & 2.850 & \citet{2010MNRAS.407..507C} \\
			\textbf{HD 15337 c} & 36.870 & -27.640 & 7.6 & 1142 & 17.180 & 2.4 & 8 & K1 V & 2.250 & \citet{2019ApJ...876L..24G} \\
			\textbf{HD 15337 b} & 36.870 & -27.640 & 7.6 & 1211 & 4.760 & 1.6 & 162 & K1 V & 2.490 & \citet{2019ApJ...876L..24G} \\
			WASP-99 b & 39.900 & -50.010 & 8.4 & 1113 & 5.750 & 11.4 & 679 & F8 & 5.260 & \citet{2014MNRAS.440.1982H} \\
			HD 17156 b & 42.440 & 71.750 & 7.1 & 1593 & 21.220 & 12.3 & 93 & G5 & 3.140 & \citet{2007ApJ...669.1336F} \\
			HR 858 d & 42.980 & -30.810 & 5.5 & 3719 & 11.230 & 2.2 & 217 & F6 V & 3.430 & \citet{2019ApJ...881L..19V} \\
			HR 858 b & 42.980 & -30.810 & 5.5 & 3305 & 3.590 & 2.1 & 990 & F6 V & 2.710 & \citet{2019ApJ...881L..19V} \\
			HR 858 c & 42.980 & -30.810 & 5.5 & 3424 & 5.970 & 1.9 & 512 & F6 V & 2.900 & \citet{2019ApJ...881L..19V} \\
			TOI-257 b & 47.520 & -50.830 & 6.5 & 3112 & 18.390 & 7.2 & 186 & F8/G0V & 6.480 & \citet{2021MNRAS.502.3704A} \\
			HD 23472 c & 55.460 & -62.770 & 7.9 & 1129 & 29.620 & 2.1 & 23 & K3.5 V & 2.940 & \citet{2019AA...622L...7T} \\
			\textbf{V1298 Tau b} & 61.330 & 20.160 & 8.7 & 1087 & 24.140 & 10.3 & 35 & K0 & 6.420 & \citet{2019AJ....158...79D} \\
			KELT-7 b & 78.300 & 33.320 & 7.7 & 1216 & 2.730 & 17.9 & 2939 & F & 3.510 & \citet{2015AJ....150...12B} \\
			\textbf{TOI-431 d} & 83.270 & -26.720 & 7.3 & 1510 & 12.460 & 3.3 & 27 & K3 & 3.240 & \citet{2021MNRAS.507.2782O} \\
			pi Men c & 84.300 & -80.460 & 4.9 & 4546 & 6.270 & 2 & 309 & G0 V & 2.950 & \citet{2018AA...619L..10G} \\
			KELT-2 A b & 92.660 & 30.960 & 7.7 & 1559 & 4.110 & 15.1 & 1617 & F8 & 5.170 & \citet{2012ApJ...756L..39B} \\
			HD 63433 c & 117.480 & 27.360 & 5.6 & 3729 & 20.550 & 2.7 & 35 & G5IV & 4.070 & \citet{2020AJ....160..179M} \\
			HD 63433 b & 117.480 & 27.360 & 5.6 & 3334 & 7.110 & 2.2 & 146 & G5IV & 3.220 & \citet{2020AJ....160..179M} \\
			HIP 41378 b & 126.620 & 10.080 & 8 & 1385 & 15.570 & 2.9 & -- & F6 & 4.860 & \citet{2016ApJ...827L..10V} \\
			55 Cnc e & 133.150 & 28.330 & 4.8 & 3478 & 0.740 & 1.9 & 2665 & G8 V & 1.580 & \citet{2004ApJ...614L..81M} \\
			MASCARA-4 b & 147.580 & -66.110 & 7.8 & 1248 & 2.820 & 17.2 & 5531 & A7 V & 3.960 & \citet{2020AA...635A..60D} \\
			HD 86226 c & 149.120 & -24.100 & 6.8 & 1882 & 3.980 & 2.2 & 449 & G1 V & 3.120 & \citet{2020AJ....160...96T} \\
			HD 89345 b & 154.670 & 10.130 & 8.1 & 1402 & 11.810 & 7.4 & 234 & G5 & 5.730 & \citet{2018MNRAS.478.4866V} \\
			KELT-11 b & 161.710 & -9.400 & 6.6 & 3045 & 4.740 & 15.1 & 1433 & G8/K0 IV & 7.140 & \citet{2017AJ....153..215P} \\
			KELT-24 b & 161.910 & 71.660 & 7.4 & 1576 & 5.550 & 14.3 & 751 & F5 & 4.300 & \citet{2019AJ....158..197R} \\
			HD 97658 b & 168.640 & 25.710 & 6.2 & 2391 & 9.490 & 2.1 & 180 & K1 V & 2.850 & \citet{2011ApJ...730...10H} \\
			HIP 57274 b & 176.170 & 30.960 & 7 & 1753 & 8.140 & 2.4 & 316 & K4 V & 3.080 & \citet{2012ApJ...745...21F} \\
			\textbf{HD 106315 c} & 183.470 & -0.390 & 8.1 & 1257 & 21.060 & 4.4 & 83 & F5 V & 4.640 & \citet{2017AJ....153..255C} \\
			HD 106315 b & 183.470 & -0.390 & 8.1 & 1151 & 9.550 & 2.4 & 240 & F5 V & 3.780 & \citet{2017AJ....153..255C} \\
			\textbf{HD 108236 f} & 186.570 & -51.360 & 8 & 1102 & 29.540 & 2 & 23 & G3 V & 3.270 & \citet{2021AA...646A.157B} \\
			\textbf{HD 108236 d} & 186.570 & -51.360 & 8 & 1189 & 14.180 & 2.5 & 68 & G3 V & 3.850 & \citet{2021AJ....161...85D} \\
			\textbf{HD 108236 e} & 186.570 & -51.360 & 8 & 1234 & 19.590 & 3.1 & 59 & G3 V & 4.200 & \citet{2021AJ....161...85D} \\
			HD 118203 b & 203.510 & 53.730 & 6.9 & 2413 & 6.130 & 13.2 & 903 & K0 & 5.710 & \citet{2006AA...446..717D} \\
			TOI-2076 b & 217.390 & 39.790 & 7.6 & 1330 & 10.360 & 3.3 & 48 & K & 3.330 & \citet{2021AJ....162...54H} \\
			HD 136352 b & 230.440 & -48.320 & 4.3 & 6813 & 11.580 & 1.7 & 112 & G4 V & 3.940 & \citet{2019AA...622A..37U} \\
			HD 136352 c & 230.440 & -48.320 & 4.3 & 6123 & 27.590 & 2.9 & 35 & G4 V & 3.250 & \citet{2019AA...622A..37U} \\
			WASP-38 b & 243.960 & 10.030 & 8.3 & 1098 & 6.870 & 13.8 & 502 & F8 V & 4.660 & \citet{2011AA...525A..54B} \\
			HAT-P-2 b & 245.150 & 41.050 & 7.8 & 1341 & 5.630 & 10.7 & 730 & F8 & 4.290 & \citet{2007ApJ...670..826B} \\
			HD 149026 b & 247.620 & 38.350 & 7.1 & 1624 & 2.880 & 8.3 & 1586 & G0 & 3.240 & \citet{2005ApJ...633..465S} \\
			\textbf{HD 152843 b} & 253.780 & 20.490 & 7.9 & 1534 & 11.630 & 3.4 & 256 & G0 & 5.530 & \citet{2021MNRAS.505.1827E} \\
			\textbf{HD 152843 c} & 253.780 & 20.490 & 7.9 & 1618 & 24.380 & 5.8 & -- & G0 & 6.360 & \citet{2021MNRAS.505.1827E} \\
			Kepler-410 A b & 283.150 & 45.140 & 8.4 & 1096 & 17.830 & 2.8 & 160 & F8 & 4.540 & \citet{2014ApJ...782...14V} \\
			Kepler-21 b & 287.360 & 38.710 & 7.2 & 1706 & 2.790 & 1.6 & 3021 & F6 IV & 3.590 & \citet{2012ApJ...746..123H} \\
			HD 183579 b & 293.290 & -54.530 & 7.5 & 1602 & 17.470 & 3.6 & 58 & G2 V & 4.360 & \citet{2021ApJ...909L...6P} \\
			KELT-20 b & 294.660 & 31.220 & 7.4 & 1377 & 3.470 & 19.5 & 4362 & A2 V & 3.580 & \citet{2017AJ....154..194L} \\
			\textbf{HAT-P-11 b} & 297.710 & 48.080 & 7.6 & 1100 & 4.890 & 4.4 & 101 & K4 & 2.360 & \citet{2010ApJ...710.1724B} \\
			\textbf{HD 189733 b} & 300.180 & 22.710 & 6.1 & 1725 & 2.220 & 12.7 & 356 & K0-2 V & 1.800 & \citet{2005AA...444L..15B} \\
			HD 191939 b & 302.030 & 66.850 & 7.6 & 1297 & 8.880 & 3.4 & 99 & G8 V & 3.080 & \citet{2020AJ....160..113B} \\
			HD 191939 c & 302.030 & 66.850 & 7.6 & 1564 & 28.580 & 3.2 & 21 & G8 V & 4.460 & \citet{2020AJ....160..113B} \\
			HD 332231 b & 306.740 & 33.740 & 7.5 & 1836 & 18.710 & 9.7 & 98 & F8 & 6.160 & \citet{2020AJ....159..241D} \\
			KELT-9 b & 307.860 & 39.940 & 7.5 & 1477 & 1.480 & 21.2 & 44900 & B9.5-A0 & 3.920 & \citet{2017Natur.546..514G} \\
			\textbf{AU Mic b} & 311.290 & -31.340 & 5.4 & 3680 & 8.460 & 4.1 & 22 & M1VE & 3.500 & \citet{2020Natur.582..497P} \\
			\textbf{AU Mic c} & 311.290 & -31.340 & 5.4 & 4220 & 18.860 & 3.2 & 7 & M1VE & 4.500 & \citet{2021AA...649A.177M} \\
			MASCARA-1 b & 317.550 & 10.740 & 7.8 & 1285 & 2.150 & 16.8 & 8111 & A8 & 4.050 & \citet{2017AA...606A..73T} \\
			HD 202772 A b & 319.700 & -26.620 & 7.2 & 2014 & 3.310 & 17.3 & 3440 & F8 & 5.630 & \citet{2019AJ....157...51W} \\
			\textbf{HD 209458 b} & 330.800 & 18.880 & 6.6 & 1861 & 3.520 & 15.6 & 768 & G0 V & 3.070 & \citet{2000ApJ...529L..41H} \\
			K2-167 b & 336.580 & -18.010 & 7.2 & 1638 & 9.980 & 2.8 & -- & F7 V & 3.270 & \citet{2018AJ....155..136M} \\
			K2-39 b & 338.370 & -9.020 & 9.1 & 1081 & 4.610 & 6.3 & 1356 & K2 & 8.180 & \citet{2016AJ....152..143V} \\
			HD 213885 b & 338.980 & -59.860 & 6.8 & 1351 & 1.010 & 1.7 & 3254 & G & 1.550 & \citet{2020MNRAS.491.2982E} \\
			HD 219134 c & 348.340 & 57.170 & 4 & 5126 & 6.760 & 1.5 & 62 & K3 V & 1.660 & \citet{2015AA...584A..72M} \\
			HD 219134 b & 348.340 & 57.170 & 4 & 3863 & 3.090 & 1.6 & 176 & K3 V & 0.940 & \citet{2015AA...584A..72M} \\
			HD 221416 b & 353.030 & -21.800 & 6.6 & 3481 & 14.280 & 9.2 & 174 & K0 IV/V & 8.490 & \citet{2019AJ....157..245H} \\
			DS Tuc A b & 354.920 & -69.200 & 7.1 & 1604 & 8.140 & 5.7 & 145 & G6 V & 3.180 & \citet{2019ApJ...880L..17N} \\
		\end{tabular}
	\end{center}
	\caption{Time-variable targets for NIGHT resulting from our SNR simulations, sorted to right ascension.}
    \end{minipage}
\end{table*}


\bsp	
\label{lastpage}
\end{document}